\documentclass[twocolumn]{aastex63}  
\usepackage{amsmath}
\usepackage{multirow}

\begin{document}

\title{COMAP Early Science: IV. Power Spectrum Methodology and Results}
\shorttitle{COMAP Early Science: Power Spectrum Results}
\shortauthors{Ihle et al.}

\correspondingauthor{H\aa vard T. Ihle}
\email{h.t.ihle@astro.uio.no}

\author[0000-0003-3420-7766]{H\aa vard T. Ihle}
\author{Jowita Borowska}
\affil{Institute of Theoretical Astrophysics, University of Oslo, P.O. Box 1029 Blindern, N-0315 Oslo, Norway}
\author[0000-0002-8214-8265]{Kieran A. Cleary}
\affil{California Institute of Technology, Pasadena, CA 91125, USA}

\author[0000-0003-2332-5281]{Hans Kristian Eriksen}
\author[0000-0001-8896-3159]{Marie K. Foss}
\affil{Institute of Theoretical Astrophysics, University of Oslo, P.O. Box 1029 Blindern, N-0315 Oslo, Norway}

\author[0000-0001-7911-5553]{Stuart E.~Harper}
\affil{Jodrell Bank Centre for Astrophysics, Alan Turing Building, Department of Physics and Astronomy, School of Natural Sciences, The University of Manchester, Oxford Road, Manchester, M13 9PL, U.K.}
\author[0000-0002-4274-9373]{Junhan Kim}
\affil{California Institute of Technology, Pasadena, CA 91125, USA}
\author{Jonas G. S. Lunde}
\affil{Institute of Theoretical Astrophysics, University of Oslo, P.O. Box 1029 Blindern, N-0315 Oslo, Norway}

\author[0000-0001-7612-2379]{Liju Philip}
\affil{Jet Propulsion Laboratory, California Institute of Technology, 4800 Oak Grove Drive, Pasadena, CA 91109, USA}
\author{Maren Rasmussen}
\author[0000-0001-5301-1377]{Nils-Ole Stutzer}
\affil{Institute of Theoretical Astrophysics, University of Oslo, P.O. Box 1029 Blindern, N-0315 Oslo, Norway}
\author[0000-0001-8526-3464]{Bade D.~Uzgil}
\affiliation{California Institute of Technology, 1200 E. California Blvd., Pasadena, CA 91125, USA}
\author[0000-0002-5437-6121]{Duncan J. Watts}
\author[0000-0003-3821-7275]{Ingunn Kathrine Wehus}
\affil{Institute of Theoretical Astrophysics, University of Oslo, P.O. Box 1029 Blindern, N-0315 Oslo, Norway}
\author[0000-0003-2358-9949 ]{J. Richard Bond}
\affil{Canadian Institute for Theoretical Astrophysics, University of Toronto, 60 St. George Street, Toronto, ON M5S 3H8, Canada}
\affiliation{Dunlap Institute for Astronomy and Astrophysics, University of Toronto, 50 St. George Street, Toronto, ON M5S 3H4, Canada}
\author[0000-0001-8382-5275]{Patrick C. Breysse}
\affil{Center for Cosmology and Particle Physics, Department of Physics, New York University, 726 Broadway, New York, NY, 10003, USA}
\author{Morgan Catha}
\affil{Owens Valley Radio Observatory, California Institute of Technology, Big Pine, CA 93513, USA}
\author[0000-0003-2358-9949]{Sarah E.~Church}
\affiliation{Kavli Institute for Particle Astrophysics and Cosmology \& Physics Department, Stanford University, Stanford, CA 94305, USA}
\author[0000-0003-2618-6504]{Dongwoo T.~Chung}
\affil{Canadian Institute for Theoretical Astrophysics, University of Toronto, 60 St. George Street, Toronto, ON M5S 3H8, Canada}
\author[0000-0002-0045-442X]{Clive Dickinson}
\affil{Jodrell Bank Centre for Astrophysics, Alan Turing Building, Department of Physics and Astronomy, School of Natural Sciences, The University of Manchester, Oxford Road, Manchester, M13 9PL, U.K.}
\author[0000-0002-5223-8315]{Delaney A.~Dunne}
\affil{California Institute of Technology, Pasadena, CA 91125, USA}
\author{Todd Gaier}
\affiliation{Jet Propulsion Laboratory, California Institute of Technology, 4800 Oak Grove Drive, Pasadena, CA 91109, USA}
\author{Joshua Ott Gundersen}
\affil{Department of Physics, University of Miami, 1320 Campo Sano Avenue, Coral Gables, FL 33146, USA}
\author[0000-0001-6159-9174]{Andrew I.~Harris}
\affiliation{Department of Astronomy, University of Maryland, College Park, MD 20742, USA}
\author{Richard Hobbs}
\affiliation{Owens Valley Radio Observatory, California Institute of Technology, Big Pine, CA 93513, USA}
\author[0000-0002-5959-1285]{James W.~Lamb}
\affil{Owens Valley Radio Observatory, California Institute of Technology, Big Pine, CA 93513, USA}
\author{Charles R.~Lawrence}
\affiliation{Jet Propulsion Laboratory, California Institute of Technology, 4800 Oak Grove Drive, Pasadena, CA 91109, USA}
\author{Norman Murray}
\affil{Canadian Institute for Theoretical Astrophysics, University of Toronto, 60 St. George Street, Toronto, ON M5S 3H8, Canada}
\author[0000-0001-9152-961X]{Anthony C.~S. Readhead}
\affil{California Institute of Technology, Pasadena, CA 91125, USA}
\author[0000-0002-8800-5740]{Hamsa Padmanabhan}
\affil{Departement de Physique Théorique, Universite de Genève, 24 Quai Ernest-Ansermet, CH-1211 Genève 4, Switzerland}
\author[0000-0001-5213-6231]{Timothy J.~Pearson}
\affil{California Institute of Technology, Pasadena, CA 91125, USA}
\author[0000-0002-1667-3897]{Thomas J. Rennie}
\affil{Jodrell Bank Centre for Astrophysics, Alan Turing Building, Department of Physics and Astronomy, School of Natural Sciences, The University of Manchester, Oxford Road, Manchester, M13 9PL, U.K.}
\author{David P. Woody}
\affil{Owens Valley Radio Observatory, California Institute of Technology, Big Pine, CA 93513, USA}
\collaboration{35}{(COMAP Collaboration)}

\begin{abstract}
We present the power spectrum methodology used for the first-season COMAP analysis, and assess the quality of the current data set. The main results are derived through the Feed-feed Pseudo-Cross-Spectrum (FPXS) method, which is a robust estimator with respect to both noise modeling errors and experimental systematics. We use effective transfer functions to take into account the effects of instrumental beam smoothing and various filter operations applied during the low-level data processing. The power spectra estimated in this way have allowed us to identify a systematic error associated with one of our two scanning strategies, believed to be due to residual ground or atmospheric contamination. We omit these data from our analysis and no longer use this scanning technique for observations. We present the power spectra from our first season of observing and demonstrate that the uncertainties are integrating as expected for uncorrelated noise, with any residual systematics suppressed to a level below the noise. Using the FPXS method, and combining data on scales $k=0.051-0.62 \,\mathrm{Mpc}^{-1}$ we estimate $P_\mathrm{CO}(k) = -2.7 \pm 1.7 \times 10^4\mu\textrm{K}^2\mathrm{Mpc}^3$, the first direct 3D constraint on the clustering component of the CO(1--0) power spectrum in the literature. 
\end{abstract}
\keywords{\href{http://astrothesaurus.org/uat/262}{CO line emission (262)}; \href{http://astrothesaurus.org/uat/336}{Cosmological evolution (336)}; \href{http://astrothesaurus.org/uat/734}{High-redshift galaxies (734)}; \href{http://astrothesaurus.org/uat/1073}{Molecular gas (1073)}; \href{http://astrothesaurus.org/uat/1338}{Radio astronomy (1338)}}

\section{Introduction}
Intensity mapping aims to map out large 3D volumes using bright emission lines as tracers of the large scale matter distribution \citep{madau1997, Battye2004, Peterson2006, Loeb2008, Kovetz2017, Kovetz2019}. One promising set of lines are the rotational transitions of the carbon monoxide (CO) molecule. CO traces cold molecular gas, and is closely linked to star formation \citep{Carilli2013}.

The CO Mapping Array Project (COMAP) is an intensity mapping experiment targeting CO.
This paper, one of a set associated with the first-season COMAP analysis, presents the methodology used to constrain the CO power spectrum with COMAP data. 
An overview of the COMAP experiment is presented by \citet{es_I}, while the COMAP instrument is described by \citet{es_II}. 

The low-level COMAP data processing pipeline is summarized by \citet{es_III}. This pipeline converts raw uncalibrated observations into three-dimensional maps, using redshifted CO line emission from distant galaxies as a tracer of the cosmic density field. Since the first-season COMAP instrument observes at frequencies between 26 and 34\,GHz, and the rotational CO(1--0) transition has a rest frequency of 115\,GHz, the current measurements trace galaxy formation at redshifts between $z=2.4$ and $3.4$, during ``the epoch of galaxy assembly". Current limits, forecasts and modeling at these redshifts is discussed in \citet{es_V}, while a future phase of COMAP targeting ``the epoch of reionization" is discussed in \citet{es_VII}. The use of this instrument for a galactic survey is presented in \citet{es_VI}.

One common and powerful quantity used to characterize the statistical properties of such three-dimensional cosmic maps is the so-called power spectrum (or the two-point correlation function), which measures the strength of fluctuations as a function of physical distance \citep[e.g.,][]{Lidz2011,Pullen2013,Li,Uzgil2019,Bernal2019,Ihle2019,Chung2019b,Gong2020,Yang2020,Keenan2021,Moradinezhad2021}. For an isotropic and Gaussian random field, this function quantifies all statistically relevant information in the original data set, but with a far smaller number of data points, and it thus represents a dramatic compression of the full data set. For non-Gaussian fields, additional information can be extracted by use of other statistics \citep{Breysse_2017, Ihle2019, breysse_etal_19, Sato2022}. Even for non-Gaussian fields, however, such as the galactic density field, the power spectrum encapsulates a large fraction of the important information, and it is therefore an efficient tool even for such fields.

However, while compressing hundreds of terabytes of raw data into a handful of power spectrum coefficients certainly makes the interpretation of the data easier in terms of theoretical comparisons, it also makes the final estimates highly sensitive to small systematic effects and instrumental noise. To guide our intuition, we note that current theories predict an intrinsic CO standard deviation per resolution element of no more than a few microkelvin \citep{Breysse2014,Li,es_V}, which is to be compared with a typical system temperature of 44\,K for the COMAP instrument; or atmospheric fluctuations of a few kelvin; or sidelobe contributions of a few millikelvin. All such effects must therefore be suppressed by many orders of magnitude in order to establish robust astrophysical constraints.

As described by \citet{es_II, es_III}, the COMAP focal plane consists of 19 different feed-horns, arranged in a hexagonal pattern, with about 12 arcmin sky separation between the closest feeds. The signal entering each feed-horn is sent through its own signal chain with its own amplifiers and digital high-resolution spectrometers. Each such signal chain is typically referred to as a ``feed". As such, the instrument has many unique features that makes it suited to this process. A few important examples include highly efficient spectroscopic rejection of common-mode signals, several semi-independent feeds, a configurable scanning strategy, and frequent usage of hardware calibrators. Still, rejection of systematic errors at the microkelvin level is highly challenging, and the current paper describes several algorithmic methods that can be applied to improve the robustness of the final results.     

The rest of the paper is organized as follows. In Section~\ref{sec:methods} we review various aspects of power spectrum methods and present our adopted COMAP power spectrum estimator, the Feed-Feed Pseudo-Cross-Spectrum (FPXS). Power spectra estimated using data from COMAP's first observing season are presented in Section~\ref{sec:results}, and we conclude in Section~\ref{sec:discussion}. 

\section{Methods}
\label{sec:methods}

We begin our discussion with an overview of the fundamental algorithms used for COMAP power spectrum estimation. For other recent examples of the use of power spectrum analysis on intensity mapping data see e.g. \cite{LOFAR_PS}, \cite{Keating2020}, and \cite{Keenan2021}. 

\subsection{Auto-spectrum analysis}
Let $m_{ijk}$ denote a three-dimensional map, and let us call each resolution element in this map a voxel. $i, j$ and $k$ are then the voxel indices.  We define the power spectrum $P(\vec k)$ of this map to be the variance of its Fourier components, $f_{\vec k}$,
\begin{equation}
    P(\vec k) = \frac{V_\mathrm{vox}}{N_\mathrm{vox}}\langle |f_{\vec k}|^2 \rangle,
\end{equation}
where $\vec k$ is the wave vector of a given Fourier mode, $V_\mathrm{vox}$ is the volume of each voxel, and $N_\mathrm{vox}$ is the total number of voxels. 

If we assume that the map is statistically isotropic, then the power spectrum will only be a function of the magnitude of the wave vector, $P(\vec k) = P(k)$. In observational cosmology we often want to distinguish between the angular directions (denoted by the $x$ and $y$ coordinates) from the line-of-sight direction (denoted by the $z$ coordinate). This is because the map typically has different properties in the different directions, for example due to instrumental beam effects or redshift space distortions \citep{Hamilton1998,Chung2019b}. It is therefore often useful to define the power spectrum in terms of parallel (line-of-sight) modes, $k_\parallel \equiv |k_z|$, and the perpendicular (angular) modes, $k_\bot \equiv \sqrt{k^2_x + k^2_y}$. We can estimate the power spectrum in a given set of $\vec k$-bins, \{$\vec k_i$\}, from a given map as
\begin{equation}
    P(\vec k_i) \approx \frac{V_\mathrm{vox}}{ N_\mathrm{vox}N_\mathrm{modes}} \sum_{j=1}^{N_\mathrm{modes}} |f_{\bf k_j}|^2\equiv P_{\vec k_i},
\end{equation}
where $P_{\vec k_i}$ is the estimated power spectrum in bin number $i$ and $N_\mathrm{modes,i}$ is the number of Fourier components with wave number ${\vec k_j} \approx \vec k_i$ (i.e. in the bin corresponding to wave number $\vec k_i$).

Assuming that foreground and systematic contributions have already been removed to negligible levels through pre-processing, the power spectrum of a cleaned line intensity map is typically modeled as a sum of a signal and noise component (assumed to be statistically independent),
\begin{equation}
    P(\vec k) = P_\mathrm{signal}(\vec k) + P_\mathrm{noise}(\vec k).
\end{equation}
If we are able to estimate the noise power spectrum through independent means, for example using a noise model or simulations, we can extract the signal power spectrum simply by subtracting the estimated noise, 
\begin{equation}
    P_\mathrm{signal}(\vec k_i) \approx P_{\vec k_i} - P^\mathrm{est}_\mathrm{noise}(\vec k_i),
\end{equation}
where $P^\mathrm{est}_\mathrm{noise}(\vec k_i)$ is the estimated noise power spectrum in bin number $i$. 

If the map consists of uniformly distributed white noise, then the noise power spectrum is independent of $\vec k$ and given by
\begin{equation}
    P_\mathrm{noise} = V_\mathrm{vox} \sigma_T^2,
\end{equation}
where $\sigma_T$ is the white noise standard deviation in each voxel (in units of kelvin). In our case, this magnitude of the white noise level is determined by the radiometer equation,
\begin{equation}
    \sigma_T = \frac{T_\mathrm{sys}}{\sqrt{\delta_\nu \tau}},
\end{equation}
where $T_\mathrm{sys}$ is the system temperature of the detector, $\delta_\nu$ is the frequency resolution of each voxel, and $\tau$ is the total time each pixel is observed. 

In addition to this instrumental noise contribution, there is an intrinsic uncertainty when estimating the signal power spectrum from a map, called sample variance, that arises from the limited number of Fourier modes in the map. Together these contributions give us the uncertainty of the power spectrum
\begin{align}
    \sigma_P &\equiv \sqrt{\langle (P_{\vec k_i} - P(\vec k_i))^2} \rangle \nonumber\\
    &\approx  \underbrace{\frac{P_\mathrm{noise}(\vec k_i)}{\sqrt{N_\mathrm{modes}}}}_{\text{Thermal noise}} + \underbrace{\frac{P_\mathrm{signal}(\vec k_i)}{\sqrt{N_\mathrm{modes}}}}_{\text{Sample variance}},
\end{align}
where $N_\mathrm{modes}$ is the number of Fourier modes in bin number $i$, and the last approximation is exact when the Fourier modes are assumed to be independent Gaussian variables. 

If the power spectrum is noise dominated, we can reduce this intrinsic uncertainty in two ways. First, we can observe for a longer time on the same area of the sky, thus decreasing the noise power spectrum contribution to the uncertainty. Alternatively, we can cover a larger sky area, and thus increase the number of measured Fourier modes. As long as we are noise dominated, a simple analysis suggests that observing a small area for a long time is more efficient for making a first detection than spreading the observations over a larger area. In a realistic situation, however,  there are several other factors that must be taken into account, including the choice of angular resolution and scanning strategy constraints, and these will typically limit how small a field it is possible to observe. 

Another source of uncertainty in estimating the signal is the accuracy of the estimated noise power spectrum model. If this model is biased or uncertain, then the associated residuals will propagate directly into the estimate of $P_\mathrm{signal}(\vec k_i)$.

\subsection{Pseudo-spectrum analysis}\label{sec:pseudo}
As described above, there are several challenges with an auto-spectrum analysis, as will be discussed both in this and the following sections. First of all, if the noise in the map is not uniform, which it generally is not, the noise power spectrum will be dominated by the parts of the map with the highest noise level. In order to address this, it is necessary to devise a method that puts more weight on the parts of the map with low noise, and less weight on the parts of the map with high noise. 

The standard method of accounting for this is through inverse noise variance weights. That is, we weigh the map, $m$, by the noise level map, $\sigma_m$ (the map given by the expected standard deviation of the white noise in each voxel), before we compute the power spectrum,
\begin{equation}
    \tilde P_{\vec k_i} = \frac{V_\mathrm{vox}}{ N_\mathrm{vox}N_\mathrm{modes}} \sum_{j=1}^{N_\mathrm{modes}} |\tilde f_{\bf k_j}|^2,
\end{equation}
where $\tilde P$ denotes the \emph{pseudo}-spectrum, and $\tilde f$ are the Fourier components of the noise weighted map,
\begin{equation}
    \tilde m \equiv w m,
\end{equation}
and 
\begin{equation} \label{eq:weightnorm}
    w \equiv \mathcal{N} \frac{1}{\sigma_m^2}.
\end{equation}
$\mathcal{N}$ is a single overall normalization constant (which we will get back to), and $\sigma_m$ is, as usual, the noise level map. 

On a general note, the term `pseudo-spectrum' typically refers to a power spectrum estimator that is computed from a biased estimator of the true sky map, and is as such itself biased; see \citealt{hivon2002}. This may be contrasted to more conventional power spectrum estimators that aim to estimate the power spectrum of the true sky signal. The statistical information content of the pseudo-spectrum and unbiased power spectrum is identical, and the main difference between the two classes of estimators concerns their ease of interpretation; while the unbiased power spectrum may be directly compared with theoretical models and other literature results, the pseudo-spectrum is experiment dependent, and typically requires simulations for proper statistical interpretation.  

In our setting, we use the pseudo-spectrum to take into account both masked voxels (by setting $\sigma_m \rightarrow \infty$ for voxels that are excluded from further analysis) and varying noise levels across the map. Both these operations lead to \emph{mode mixing}, i.e., different signal Fourier modes are mixed together, and the estimated signal pseudo-spectrum is therefore a distorted version of the true signal power spectrum. However, since we know exactly how the signal map has been distorted, we can, at least in principle, calculate the exact mode mixing matrix that is needed to reconstruct the mode mixing and obtain an unbiased signal spectrum from the pseudo-spectrum \citep{hivon2002}. How feasible this is for a specific case depends on the details of the map dimensions and computational resources. For more details on mode mixing, see Appendix~\ref{app:mode_mixing}. 
 
Although mode mixing does complicate the physical interpretation of the pseudo-spectrum, there are several ways of dealing with this without having to calculate and invert the full mode mixing matrix. First of all, if the analysis involves comparisons with signal simulations, then one may simply apply the same weight matrix to each simulation, making the observed and simulated power spectra statistically compatible. Second, if the level of mode mixing is modest, then the pseudo-spectrum may be an adequate estimator for the signal power spectrum for a given application, especially on smaller scales. This typically holds particularly well for noise-dominated applications, for which a single power estimate covering a large range in $k$ is desired; in that case, the mode mixing often has minimal effect on the estimates, and the pseudo-spectrum often is a perfectly valid estimate in its own right. The accuracy of this approximation must be assessed for each use case. 

In cases for which the pseudo-spectrum is intended to be used as a direct estimator, it is necessary to choose a value for the normalization factor $\mathcal{N}$ in Equation~\ref{eq:weightnorm}. Establishing the \emph{formally correct} value for this normalization is not entirely well defined, as you are essentially trying to approximate the effect of an entire matrix with a single number (see Appendix~\ref{app:mode_mixing} for more details). However, we can make a simple and fairly reasonable choice as follows
\begin{equation}
    \mathcal{N} = \frac{1}{\sqrt{\left\langle \frac{1}{\sigma_m^4}\right\rangle}},
\end{equation}
where $\langle\rangle$ denotes average over the whole map. To make the results easier to interpret, we therefore apply this normalization to all results shown in this paper. For analyses that employ the full mode-mixing matrix, or in which the pseudo-spectrum is compared directly to simulations, this normalization is completely irrelevant. 

\begin{figure}[t]
	\begin{center}
        \includegraphics[width=0.48\textwidth]{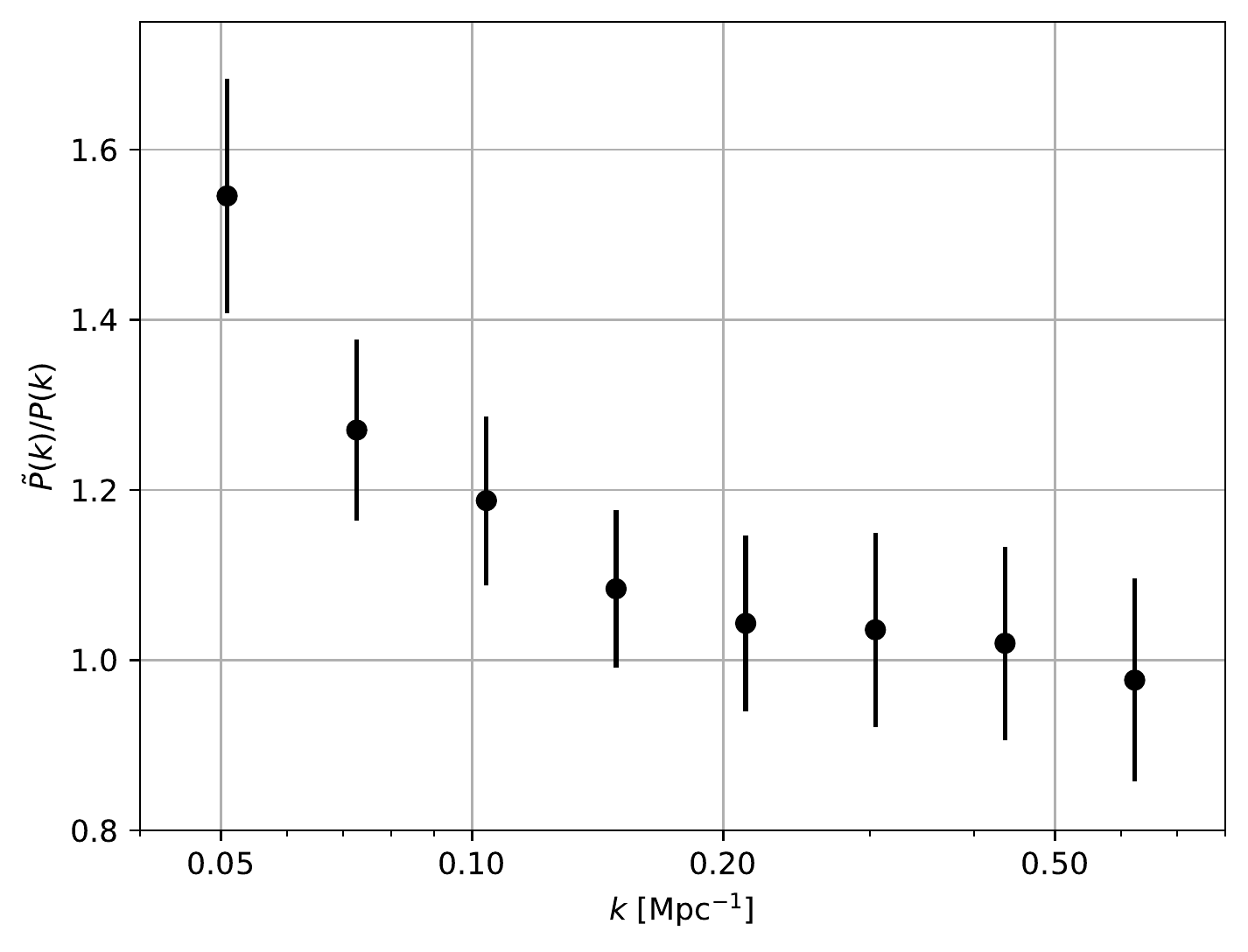}   
		\vskip -1em
		\caption{Ratio of the signal pseudo-spectrum to the signal auto-spectrum based on ten signal realizations.\label{fig:mode_mix}}
	\end{center}
\end{figure}
To roughly estimate the expected level of mode mixing we calculate the ratio of the pseudo-spectrum and the auto-spectrum for ten signal realization maps. Figure~\ref{fig:mode_mix} shows the mean and standard deviation of the mode mixing in each of the main power spectrum bins. Overall, we see that at the scales where we have most of our sensitivity, the effect of mode mixing is fairly modest, typically in the 5--30\% range. Thus, even at face value, the pseudo-spectrum does provide a reasonable order-of-magnitude estimate of the true power spectrum, even if it may not be appropriate for precision analysis. We also note that these results suggest that, if anything, an upper limit obtained by interpreting the pseudo-spectrum at face value will be a slightly weaker (i.e. more conservative) upper limit than we would get by accounting for the mode mixing. 

We leave it for future work to estimate the mode mixing matrix and undo the mode-mixing bias in the pseudo-spectra. For the rest of this paper we will interpret the pseudo-spectra at face value.  

\subsection{Cross-spectrum analysis}
A general challenge when using either the auto- or pseudo-spectrum is that highly accurate estimates of the noise contribution are required to estimate the signal power spectrum. In many cases this can be very challenging, and any systematic error will directly bias the final signal estimate. 

One way to avoid this complication is to use the so-called cross-spectrum, $C(\vec k)$. While the power spectrum quantifies the variance of the Fourier components of a single map, the cross-spectrum quantifies the covariance between the Fourier modes of two different maps,
\begin{align}
    C(\vec k) &= \frac{V_\mathrm{vox}}{N_\mathrm{vox}}\Big\langle \mathrm{Re}\{f^{*}_{1\vec k} f^{}_{2\vec k}\} \Big\rangle \nonumber\\ 
    &\approx \frac{V_\mathrm{vox}}{ N_\mathrm{vox}N_\mathrm{modes}} \sum_{j=1}^{N_\mathrm{modes}} \mathrm{Re}\{f^{*}_{1\vec k_j} f^{}_{2\vec k_j}\}\equiv C_{\vec k_i}.
\end{align}
Here $\mathrm{Re}\{\}$ denotes the real part of a complex number, and $f_1$ and $f_2$ are the Fourier components of two maps $m_1$ and $m_2$. 

Clearly, if $m_1$ and $m_2$ are identical, then the cross-spectrum is equivalent to the auto-spectrum. The advantage of the cross-spectrum, however, is that, if the maps $m_1$ and $m_2$ are made from different data, then the noise contributions are independent, and they do not contribute to the mean of the cross-spectrum, but only to its variance. Therefore, it is not necessary to estimate and subtract the noise power spectrum to obtain an unbiased signal estimate, but rather
\begin{equation}
    \langle C_{\vec k_i}\rangle = P_\mathrm{signal}(\vec k_i).
\end{equation}
Of course, a proper noise estimate is still necessary for uncertainty estimation, but the requirements on this are typically far less stringent than for the estimator mean.  

Although the cross-spectrum significantly reduces the precision needed when estimating the noise power spectrum, we do pay a price in the form of somewhat lower intrinsic sensitivity. For instance, when splitting the data into two independent parts, and cross-correlating these, we do lose a factor of at least $\sqrt{2}$ from the fact that we do not exploit the auto-correlations within each data set separately. Fortunately, this problem can be remedied by splitting the data into more independent maps, and averaging the cross-spectra of all possible combinations. A lower limit on the cross-spectrum sensitivity is given by
\begin{equation}
    \sigma_C^{N_\mathrm{split}} \geq \sqrt{\frac{1}{1 - 1/N_\mathrm{split}}} \sigma_P,
\end{equation}
where $N_\mathrm{split}$ is the number of different map splits, and $\sigma_P$ is the optimal sensitivity of the auto-spectrum derived from the full data set. 

The cross-spectrum has some other very important advantages with respect to the auto-spectrum as well. As discussed in the introduction, one of the major challenges for an experiment like COMAP, in which we have to integrate down the noise by several orders of magnitude in order to measure a small signal, are systematic errors. However, since the cross-spectrum may only be biased by structures common to the two maps, one can try to ensure that any known systematic effect contribute independently to the two maps. In that case, the systematic effects will not bias the signal estimate. Combining this insight with splitting the data into multiple parts allows us to design a power spectrum statistic that is far more robust to systematics than the auto-spectrum. 

We define a pseudo-cross-spectrum in an analogous manner as for the pseudo-auto-spectrum. The only subtlety is that we make sure to apply the same weight map, $w$, for both maps. Explicitly, we adopt the following weight map,
\begin{equation}
    w_{1,2} \propto \frac{1}{\sigma_{m_1}\sigma_{m_2}},
\end{equation}
for both $m_1$ and $m_2$ when calculating the pseudo-cross-spectrum, $\tilde C_{\vec k_i}$. 

\subsection{The Feed-feed Pseudo-Cross-Spectrum}
The idea of the Feed-feed Pseudo-Cross-Spectrum (FPXS) method is to combine all the insights from the preceding sections to construct a single statistic for the CO signal that has a high intrinsic sensitivity, uses proper noise weighting, and that is robust against instrumental and other systematic errors. In that respect, we first note that the COMAP focal plane consists of 19 feeds, each with its own amplifiers and detectors. Furthermore, many systematic errors are particular to each feed, due to different passbands, amplifiers, cables, beams etc. We may therefore split the data according to feeds (i.e., make one map per feed), and then compute cross-spectra of all different feed combinations, while never correlating two maps from the same feed. 

Second, we also note that one of the most troublesome systematic errors for COMAP is ground pickup. This is mainly because the ground contamination correlates with the pointing, and it therefore does not average down the same way as any systematic error that is random in the time-domain (and hence independent in different observations). We can make ourselves as robust as possible to any residual ground signal in our map by also splitting the data by the elevation of the observations, so that we never take the cross-spectrum of two different datasets taken at the same elevation\footnote{The ground contamination also depends on azimuth, but since most of the problematic ground contamination happens at the highest or lowest elevations, it is most natural to divide the data according to elevation.}.

With these considerations in mind, we define the following procedure for calculating the FPXS:
\begin{enumerate}
    \item We split the data into disjoint sets sorted according to elevation. For simplicity we assume for now that we split the data into two sets, $A$ and $B$, where $A$ contains all the observations taken at elevations below the median elevation, and $B$ contains all the observations from the higher elevations. We can easily generalize this to a case where we split the data into more than two sets. 
    \item For each set, $A$ and $B$, we generate maps for each of the 19 feeds. We denote the different maps according to dataset and feed, such that $A_{13}$ indicates the map that combines all data from dataset $A$ for feed number 13. 
    \item We then calculate the pseudo-cross-spectrum, $\tilde C^{ij}_{\vec k_i}$ for all different map combinations of $A_i$ and $B_j$ where $i \neq j$.
    \item Next, we compute the average pseudo-cross-spectrum, $\tilde C_{\vec k_i}$, by noise weighting all different cross-spectra,
    \begin{equation}
        \tilde C^\mathrm{FPXS}_{\vec k_i} = \left(\sum_{i\ne j} \frac{1}{\sigma^2_{\tilde C^{ij}_{\vec k_i}}}\right)^{-1}  \sum_{i\ne j} \frac{\tilde C^{ij}_{\vec k_i}}{\sigma^2_{\tilde C^{ij}_{\vec k_i}}}.
    \end{equation}
    Here $\sigma_{\tilde C^{ij}_{\vec k_i}}$ is the uncertainty (standard deviation) in $\vec k$ bin number $i$ of the pseudo-cross-spectrum of the maps $A_i$ and $B_j$, and the sum is over all combinations of $i$ and $j$ except the cases where $i = j$. Under the naive assumption that all cross-spectra are independent, the uncertainty of the combined cross-spectrum is given by
    \begin{equation}
        \sigma_{\tilde C^\mathrm{FPXS}_{\vec k_i}} = \left(\sum_{i\ne j} \frac{1}{\sigma^2_{\tilde C^{ij}_{\vec k_i}}}\right)^{-1/2}.
    \end{equation}
\end{enumerate}
The data can of course be split in other ways, to make ourselves less susceptible to other systematic effects, but we have found that using the feeds and elevation splits yields good results for the current dataset.

\begin{figure}[t]
	\begin{center}
	    \includegraphics[width=0.48\textwidth]{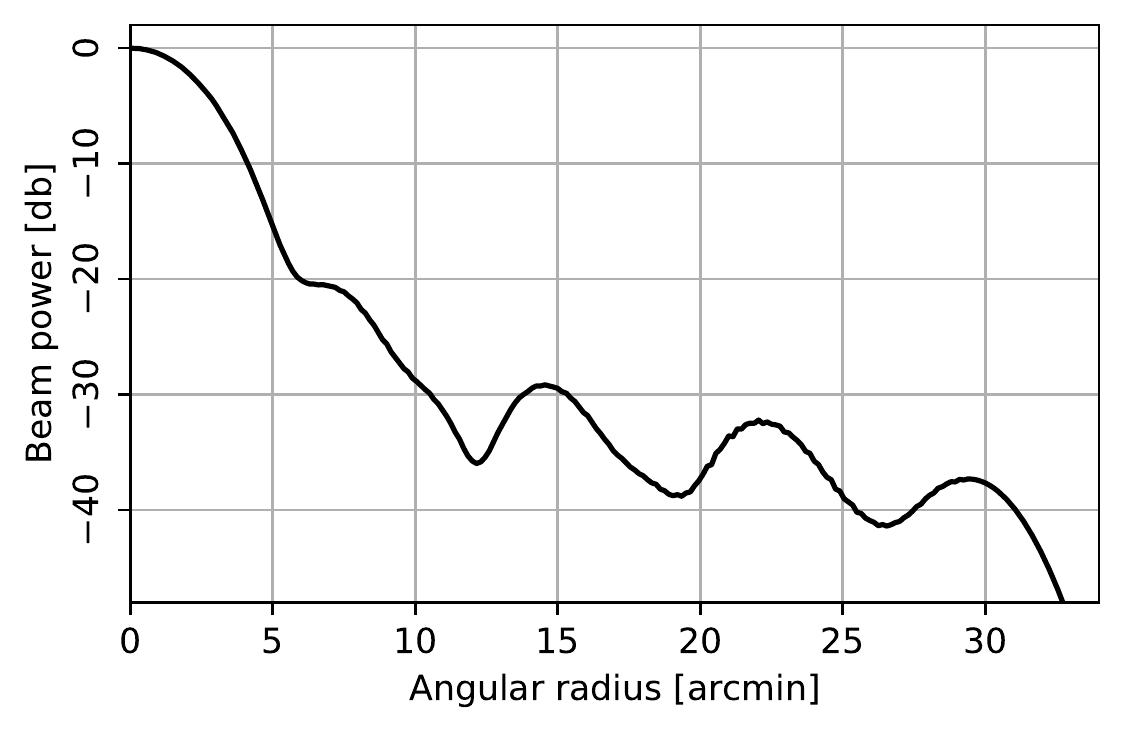}
	    \vskip -0.5em
		\includegraphics[width=0.48\textwidth]{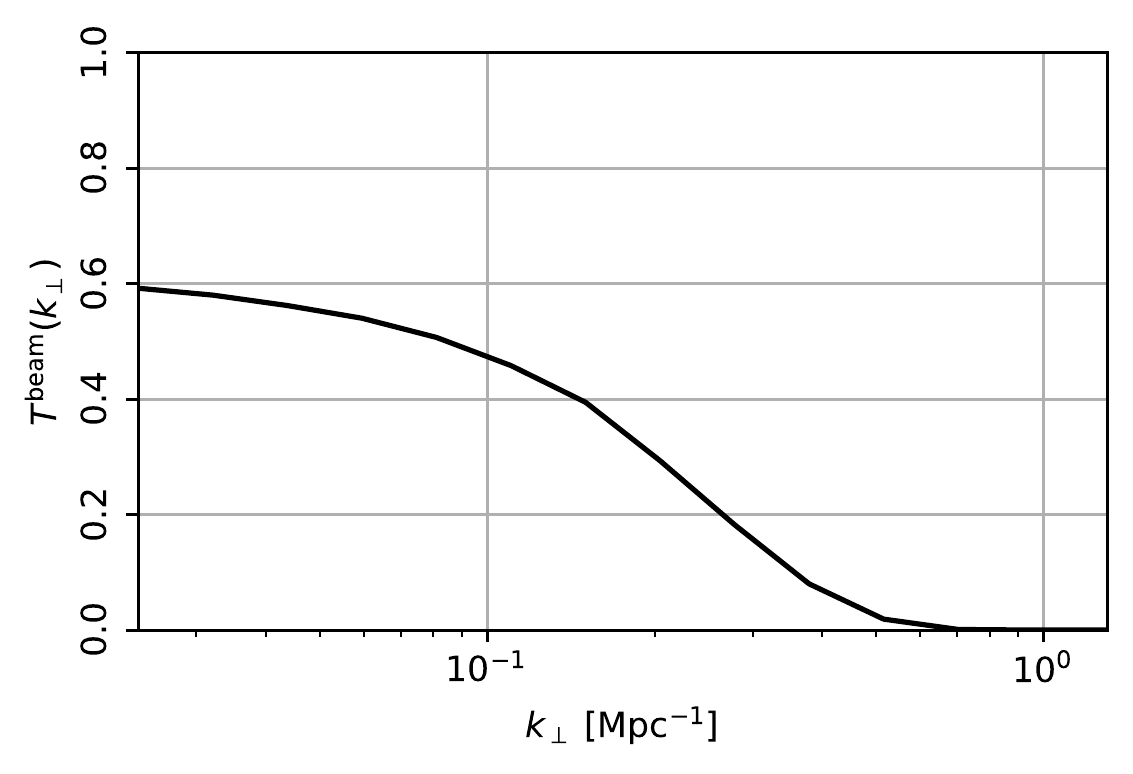}
		\vskip -1em
		\caption{Top: Radially symmetric instrumental beam model. Bottom: Resulting beam transfer function, after taking into account the main beam efficiency.\label{fig:beam_tf}}
	\end{center}
\end{figure}

\subsection{White noise simulations}
Until now we have not discussed how to estimate the noise power spectrum and the corresponding noise uncertainty of the power spectrum. In general, estimating the noise power spectrum precisely is very difficult, since one needs to take into account not only the intrinsic white noise level of the data, but also the effect of the different filtering procedures in the low-level data analysis, as well as any correlated noise contribution. 

Since we use a cross-spectrum method, however, the noise spectrum is only used to estimate the uncertainties of the power spectrum, not its mean level, and the requirements on the absolute noise spectrum are therefore somewhat relaxed. Explicitly, if we make an error of a few percent in our noise estimate, we will not bias the estimated signal spectrum, only misestimate the error bars by a few percent. While clearly not ideal, this is usually not critical, considering all the other simplifying assumptions introduced in the analysis. On the other hand, if we had adopted an auto-spectrum method, an error of a few percent on the noise power spectrum could easily have rendered our signal estimate unusable, even in the case of very high intrinsic sensitivity. 

For this reason, we therefore adopt a simple approach to noise power spectrum estimation: We assume that the noise in the maps is uncorrelated white noise, and generate noise simulations, $m_i$, by drawing random samples in each voxel from a Gaussian distribution with zero mean and a standard deviation given by the value of the noise level map, $\sigma_m$. We then estimate error bars by generating a large number of noise simulation maps, calculating the power spectrum from each, and finally taking the standard deviation in each $\vec k$ bin of interest. This gives us uncertainties on the noise contribution to each power spectrum bin, but neglects the intrinsic uncertainty in the signal power spectrum itself. However, as we are still completely noise dominated, this intrinsic uncertainty of the signal spectrum should be negligible.

\subsection{Transfer functions}
Until now we have assumed that the sky maps produced by the low-level analysis pipeline are unbiased. For multiple reasons, this is not the case. First of all, the instrument does not have infinite resolution, and the instrumental beam will therefore smooth out the signal on small angular scales. The same effect happens due to the finite spectral resolution of the instrument in the frequency dimension. Secondly, the various filters and mapmaking procedures in the analysis pipeline generally remove some of the signal, mostly on larger angular and spectral scales. In the following, we take these effects into account through so-called transfer functions. These are functions in the $k_\parallel$-$k_\bot$ plane that quantify the fraction of the signal power that is retained in each $\vec k$-bin, and allow us to establish unbiased estimates of the power spectrum from biased sky maps. 

In general a transfer function, $T(\vec k)$, is defined through the following relation,
\begin{equation}\label{eq:tf_def}
    \langle P_{\vec k} \rangle =  T(\vec k) P_\mathrm{signal}(\vec k) + P_\mathrm{noise}(\vec k),
\end{equation}
where $P_{\vec k}$ is the power spectrum calculated from the final map and $P_\mathrm{signal}(\vec k)$ is the actual physical signal power spectrum. We decompose the full transfer function into different parts, and derive each separately. We then multiply the transfer functions together to get the full transfer function. 

In writing down Equation \ref{eq:tf_def}, with a transfer function, $T(\vec k)$, which is not a function of the signal, we have implicitly assumed that the effects we are accounting for using the transfer function are linear, do not depend on the properties of the signal, and can therefore be estimated using any signal model. While this is a good approximation in many cases (e.g.\ the beam effect is purely linear and most of the low-level filters are linear, assuming the same scanning pattern), it is only an approximation. However, even any residual theoretical dependence on the input signal will typically be small in practice since the signal power spectrum in any reasonable model is very smooth and has no sharp features. The most important effect to get right when estimating the transfer function is to get the right noise distribution and scanning pattern. This is even more important since we are working with pseudo-spectra, where the noise level will affect the weighting and the mode mixing. That is why, even though it is costly to produce simulated data, we use about 63 hours of simulated data (thus ensuring a realistic scanning pattern and noise distribution) when we estimate the pipeline transfer function. Since we are anyway using the pseudo-spectrum, and not accounting for the mode mixing, we are already accepting errors of the order of 10\,\%, which puts less stringent constraints on the precision of the rest of our procedures.

\begin{figure}[t]
	\begin{center}
		\includegraphics[width=0.49\textwidth]{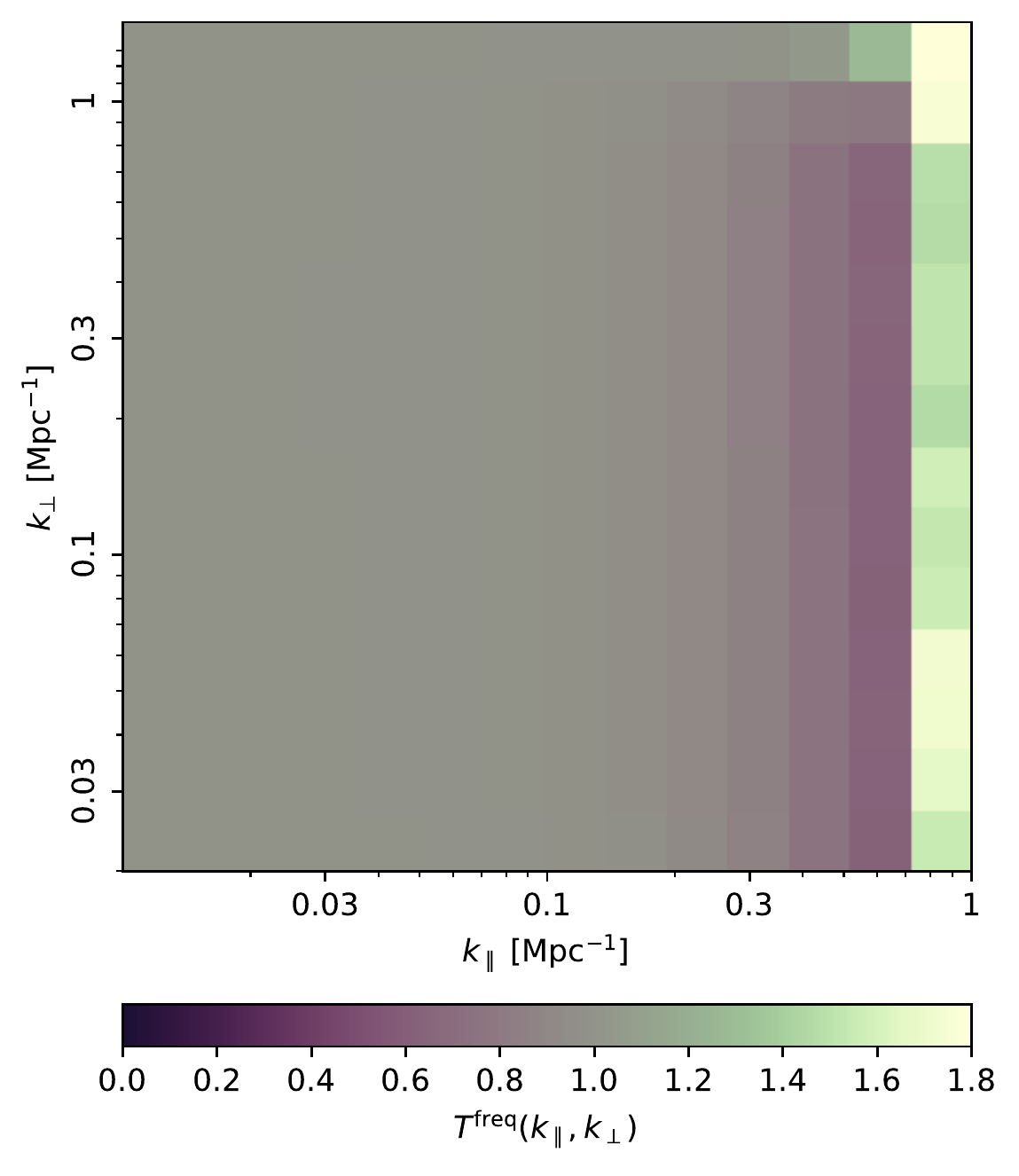}
		\caption{Frequency binning transfer function. \label{fig:pixwin_tf}}
	\end{center}
\end{figure}

\begin{figure}[t]
	\begin{center}
		\includegraphics[width=0.49\textwidth]{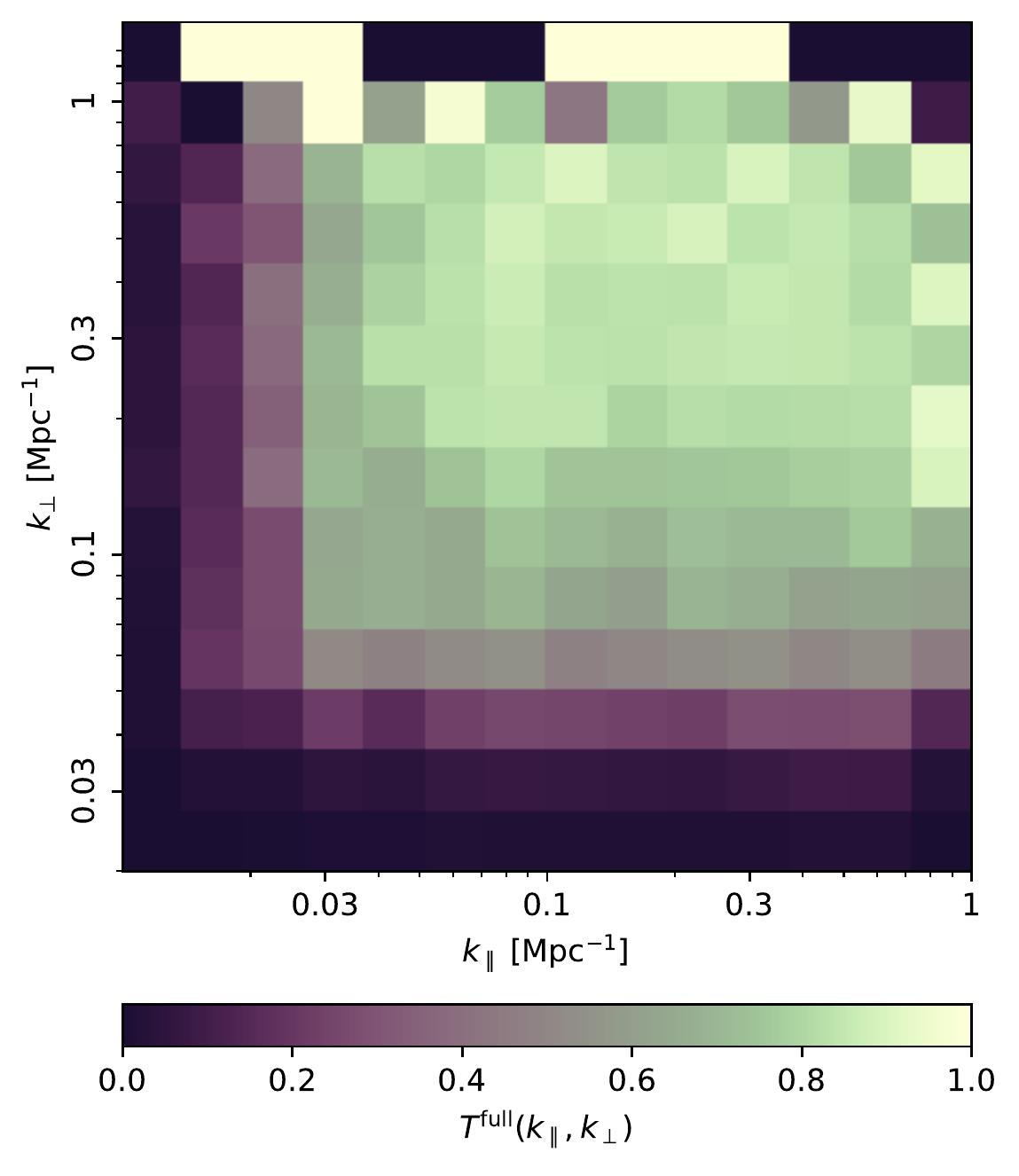}
		\vskip -1em
		\caption{Pipeline transfer functions for the cylindrically averaged power spectrum for constant elevation scans. This transfer function is based on a single signal realization and roughly three hours of data. \label{fig:TF_2d_CES}}
	\end{center}
\end{figure}

\subsubsection{Instrumental beam transfer function}\label{sec:beam_tf}
Due to the finite resolution of the instrument, we cannot measure the cosmological signal on the smallest angular scales. In order to take this effect into account we introduce a beam transfer function. For now, we assume the beam to be both achromatic (i.e., constant in frequency) and azimuthally symmetric. We construct an azimuthally symmetric beam model by averaging the full 2D (Az-El) beam model \citep{es_II} and inserting an exponential cutoff at around 30 arcmin. 

The ambient load calibration discussed in \citet{es_III} measures all of the power entering the feed horns, including that which comes from the ground and all of the sky above the horizon. However, any power on scales larger than the modes we are sensitive to is essentially lost. To get a proper, scale dependent, calibration, our beam model is normalized using observations of Jupiter and TauA which show that 72\,\% of the power is in the central 6.4 arcmin of the beam \citep{es_VI}.

In addition, by including the beam model out to about 30~arcmin we take into account the extra roughly 10~\% of power that is retained at larger angular scales. We could include the beam model further out, but we are already hitting diminishing returns, so not much more would be gained. 

Our (unnormalized) beam model can be seen in the top panel of Figure~\ref{fig:beam_tf}. The corresponding beam transfer function is estimated using signal-only simulations. That is, we generate a large number of 3D signal realizations and convolve our azimuthally symmetric beam model with the angular dimensions of the map. We then calculate the power spectrum of each of the signal realization maps with and without beam smoothing. The estimated transfer function is given as the ratio of the average of these, 
\begin{equation}
    T^\mathrm{beam}(\vec{k}) \approx \frac{\left\langle P^\textrm{signal,beam}_{\vec k}\right\rangle}{\left\langle P^\textrm{signal}_{\vec k}\right\rangle},
\end{equation}
where $P^\textrm{signal,beam}_{\vec k}$ is the power spectrum calculated from a beam smoothed signal realization map and $P^\textrm{signal}_{\vec k}$ is the power spectrum calculated from non-smoothed one.

\begin{figure}[t]
	\begin{center}
		\includegraphics[width=0.49\textwidth]{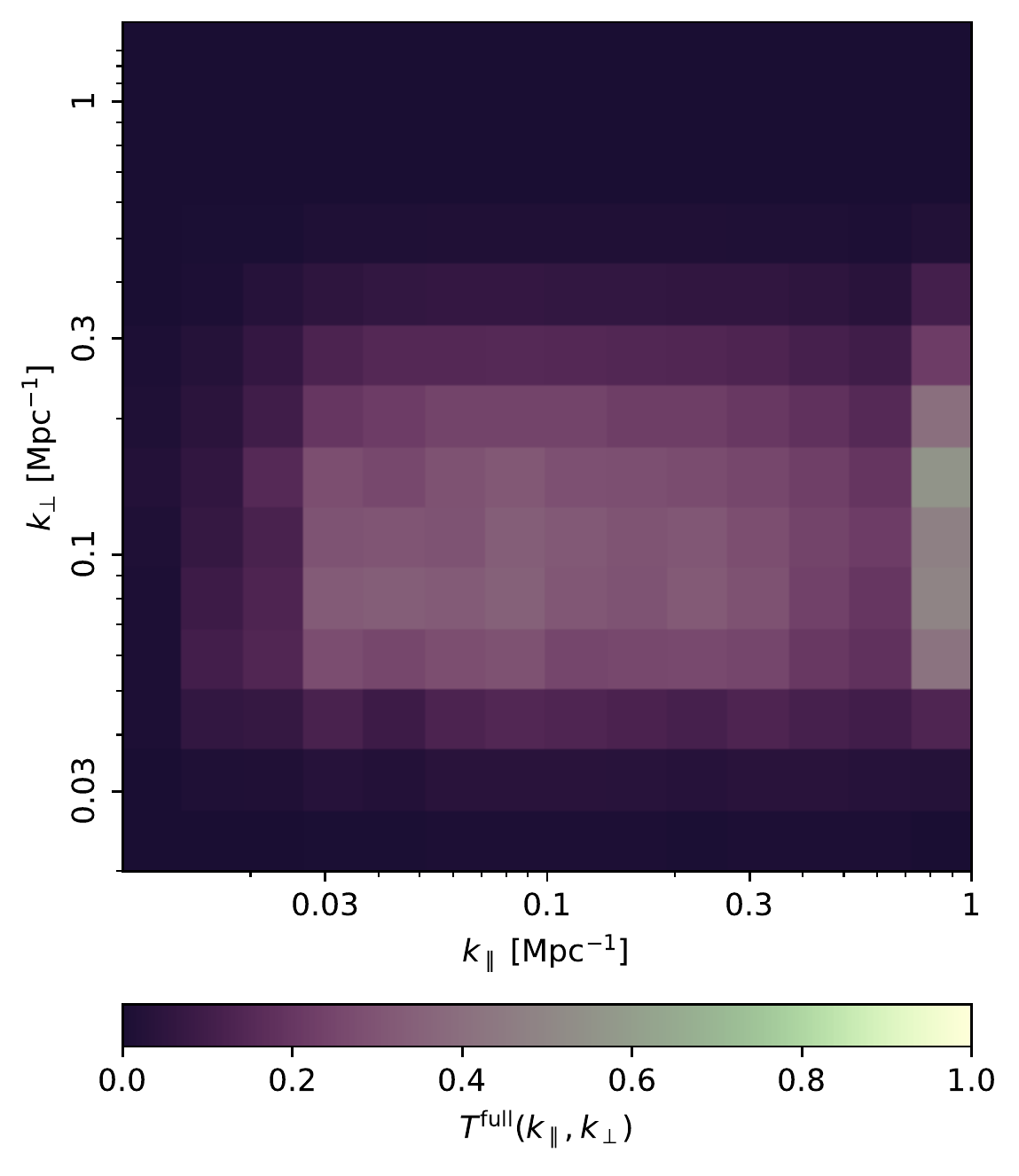}
		\caption{Full transfer function. \label{fig:full_tf}}
	\end{center}
\end{figure}

Figure~\ref{fig:beam_tf} shows the beam transfer function derived using 100 signal simulations. We see that the beam smoothing suppresses power on small angular scales, corresponding to the main beam FWHM of about 4.5 arcmin. We also see that although we lose sensitivity from our main beam efficiency, we retain some of this power on larger scales by making use of a beam model up to around 30 arcmin. 

\subsubsection{Frequency resolution transfer function}
Our current analysis, for simplicity and computational efficiency, uses fairly wide bins in frequency, of 31.25 MHz. This can be compared to intrinsic CO linewidths of order 30 MHz \citep{Chung2021a}, which will give the smallest scales present in the CO signal we are trying to observe. Once we are ready to claim a detection, we will increase the frequency resolution by at least a factor of two to get slightly more sensitivity to the small-scale CO signal, but for now this is not a high priority. We will, however, take into account the bias induced by the current bin size. Often such effects are taken into account by applying an analytic pixel window function, but this is not sufficient here since the presence of structure on scales smaller than our bins means that some of this power can be aliased into our power spectrum. 
As the effect depends on the small scale structure of the signal, there is no model-independent way to take into account this effect, and we will have to use simulations. 

We estimate the frequency binning transfer function, $T^\mathrm{freq}(\vec k)$, by comparing power spectra of simulated signal (using the default model in \citealt{es_V}) on a high resolution frequency grid to power spectra of simulated signal on our current frequency grid, both binned in our current $k_\parallel \times k_\bot$ bins. The transfer function derived using 50 such signal simulations are shown in Figure~\ref{fig:pixwin_tf}. We see a decrease in power towards smaller line of sight scales, but with an increase in the final bin, the latter of which we believe is the effect of aliasing of smaller scale structure into this bin. 

\subsubsection{Pipeline transfer function}\label{subsec:PipelineTF}
Each step of the analysis pipeline, including low-level filtering, calibration and mapmaking, affects how much of the true sky signal is present in the final maps and power spectra. We estimate the transfer function of these operations by processing the sum of the raw data and a known signal-only time-ordered simulation through the analysis pipeline, following the exact same procedure as for the raw data alone. The pipeline transfer function may then be estimated as 
\begin{equation}
    T^\mathrm{pipeline}(\vec{k}) \approx \left\langle\frac{P^\textrm{full}_{\vec k} - P^\textrm{noise}_{\vec k}}{P^\textrm{signal}_{\vec k}}\right\rangle,
\end{equation}
where $P^\textrm{full}_{\vec k}$ is the power spectrum calculated from the maps derived from the raw data with added signal, $P^\textrm{noise}_{\vec k}$ is the spectrum derived from the same data but without the added signal, and $P^\textrm{signal}_{\vec k}$ is the power spectrum derived from the raw signal simulation that was added to the raw data.

In Figure~\ref{fig:TF_2d_CES} the 2D binned pipeline transfer function for the CES data is shown. The transfer function peaks at intermediate $k$'s, with efficiencies of $\sim 0.8-0.85$ around the peak region. We see that we lose the largest scales both in the angular and the line of sight directions. This is due to the various filters applied to the time-ordered data to remove correlated noise and systematics, in addition to the effects of the scanning strategy.  For more details see \citet{es_III}.

\subsubsection{Unbiased signal estimate}
Figure \ref{fig:full_tf} shows the full transfer function combining all the effects discussed above. 
Correcting the FPXS with the above transfer function, we can establish an unbiased estimate of the signal pseudo-spectrum,
\begin{equation}
    \tilde P_\mathrm{signal}(\vec k) \approx \tilde C_{\vec k} \equiv\frac{\tilde C^\mathrm{FPXS}_{\vec k}}{\tilde T^\mathrm{full}_{\vec k}},
\end{equation}
where $\tilde P_\mathrm{signal}(\vec k)$ is the signal pseudo-spectrum and $\tilde T^\mathrm{full}_{\vec k} = \tilde T^\mathrm{beam}_{\vec k} \tilde T^\mathrm{freq}_{\vec k} \tilde T^\mathrm{pipeline}_{\vec k}$ is the full estimated transfer function for the pseudo-spectrum. The uncertainty of this signal estimate is given by
\begin{equation}
    \sigma_{\tilde P_\mathrm{signal}(\vec k)} = \sigma_{\tilde C_{\vec k}} \equiv \frac{\sigma_{\tilde C^\mathrm{FPXS}_{\vec k}}}{\tilde T^\mathrm{full}_{\vec k}} .
\end{equation}

\subsubsection{Spherical averaging}
Due to the transfer function, different $\vec k$-modes corresponding to the same $k = |\vec{k}|$ bin have very different sensitivity. In order to get the best result, we need to take this into account when we calculate the spherically averaged power spectrum. As before we use inverse variance noise weighting to achieve this, giving us the following estimate for the unbiased spherically averaged pseudo-cross-spectrum
\begin{equation}
    \tilde C_{k_i} \equiv \frac{1}{\sum_{|\vec{k}| \in k_i} w_{\vec{k}}} \sum_{|\vec{k}| \in k_i} w_{\vec{k}}\tilde C_{\vec{k}},
\end{equation}
where $w_{\vec{k}} \equiv 1/\sigma^2_{\tilde C_{\vec k}}$ and where $k_i$ denotes the $i$'th $k$ bin. 

For simplicity, we only do the spherical average of the cross-spectra that have already been cylindrically averaged and binned. This means that we use the bin centers of the $k_\parallel \times k_\bot$ bins to represent all the modes in the bin, which means that a few $\vec{k}$ modes get shifted back or forth by one bin in the spherically averaged cross-spectrum. Since there are no sharp features in the signal power spectrum this bias is modest, and not very important for the first-season analysis.

\section{Power spectrum results}
\label{sec:results}
As described in \citet{es_III}, after the COMAP time-ordered data have been filtered and calibrated, and bad observations have been removed, the cleaned data set is compressed into a set of 3D maps. We make separate maps for the Lissajous scans and the CES, since these tend to have different systematics and statistical properties.

\subsection{FPXS results}
We estimate the cross-spectrum separately for the Lissajous and CES data, for each of the three CO fields that we have observed \citep{es_III}. Since we found clear excess power in the Lissajous spectra we do not include them in the main results, and we will here focus on the CES data. Power spectrum results for the Lissajous data are presented in Appendix~\ref{app:liss}. 

We split the data in two parts according to the elevation of the observations, and use the FPXS method on these two sets of feed maps in order to minimize systematics. We also calculate a $\chi^2$ statistic for each of the 16 $\times$ 15 different feed-feed cross-spectra\footnote{As discussed in \citet{es_III}, all data from feeds 4, 6 and 7 are rejected at an earlier stage of data selection. This leaves the data from 16 out of the full 19 feeds.}, $\tilde C^{ij}_{\vec k_i}$. 

Based on these $\chi^2$ statistics, denoted $\chi^2_{C(\vec k)}$, we decided to reject all the spectra involving feed 8, since they showed very clear excesses in almost all spectra. This reduced the amount of data by 12.5 \% for all fields. We also saw clear structure in several of the spectra involving the low elevation data from feeds 16 and 17 in the Field 1 results. This led us to remove all spectra involving these feeds from the low elevation dataset for Field 1, thereby increasing the data loss to 24.2 \% for this field. 
In addition to the spectra that were removed by hand we also reject all spectra with more than a 5 sigma excess in $\chi^2_{C(\vec k)}$, before we calculate the FPXS mean spectrum. 

In the automatic 5$\sigma$ cut, the fraction of remaining spectra that were removed for the CES data was given by 1/182 for Field 1, 159/210 for Field 2 and 65/210 for Field 3. As discussed in \citet{es_III}, the fact that such a large fraction of the data is removed at this stage (especially for Fields 2 and 3) suggests that large improvements in sensitivity can be achieved in the future if we can identify the data affected by systematic errors at an earlier stage of the pipeline. 

\begin{figure}[t]
	\begin{center}
		\includegraphics[width=0.49\textwidth]{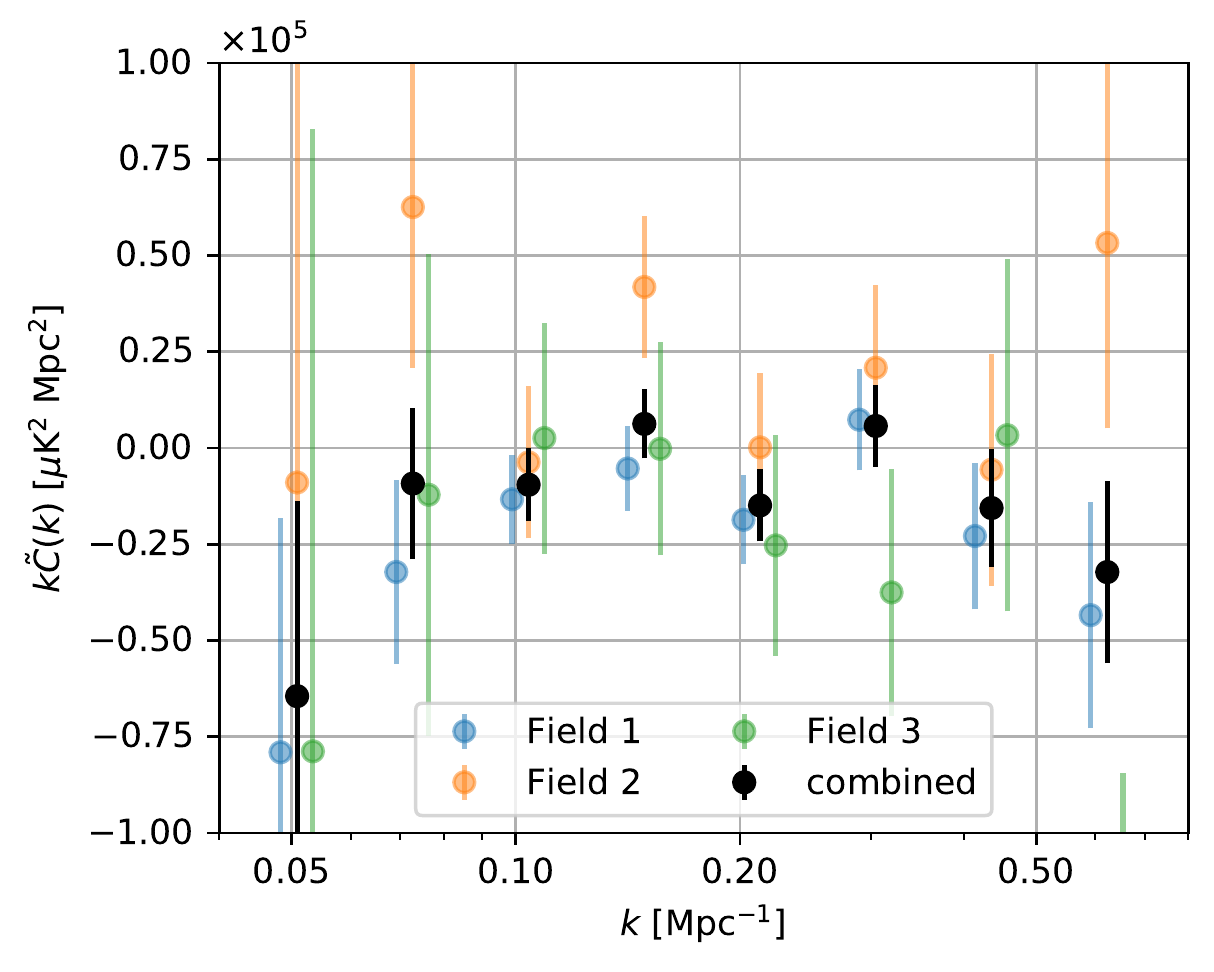}
		\caption{Spherically averaged mean pseudo-cross-spectra for CES observations of Field 1 (blue), Field 2 (orange) and Field 3 (green). These spectra were generated from all the accepted data using the FPXS cross-spectrum statistic. In addition the full transfer function has been applied, to de-bias the signal estimate. Data points from the different fields are offset slightly in $k$ from their actual values to make them easier to distinguish. \label{fig:FPXS_1d_CES}}
	\end{center}
\end{figure}

The resulting spherically averaged pseudo-cross-spectra are shown in Figure~\ref{fig:FPXS_1d_CES}. We see that the results for the CES data appear largely flat, with fluctuations that are consistent with our white noise estimate. This demonstrates that we are in fact averaging down the noise as expected for uncorrelated noise, and that the various potential systematic errors are suppressed to a level below the noise. At this point in the COMAP survey, this is a key outcome, given our fiducial theoretical model predicts a signal on the order of $kP_\mathrm{CO}\sim 10^3$~$\mu$K${}^2$~Mpc${}^2$ at our target redshift \citep{es_V}, well below the noise level shown here. This signal estimate is highly uncertain, however, and as discussed in \citet{es_V} these data already rule out some of the most optimistic models. 

Combining these datapoints into a single measurement of the average CO power spectrum over the range $k=0.051-0.62 \,\mathrm{Mpc}^{-1}$ we get 
\begin{equation}
    P_\mathrm{CO}(k) = -2.7 \pm 1.7 \times 10^4 \mu\textrm{K}^2\mathrm{Mpc}^3.
\end{equation}
This estimate is based on the pseudo-spectrum, and, as discussed in Section~\ref{sec:pseudo}, it is a somewhat biased estimate of the signal, but should be a conservative estimate if used as an upper limit, as we will do in \citealp{es_V}. This is the first direct 3D constraint on the clustering component of the CO(1--0) power spectrum in the literature.

\begin{figure*}[t]
	\begin{center}
		\includegraphics[width=1.0\textwidth]{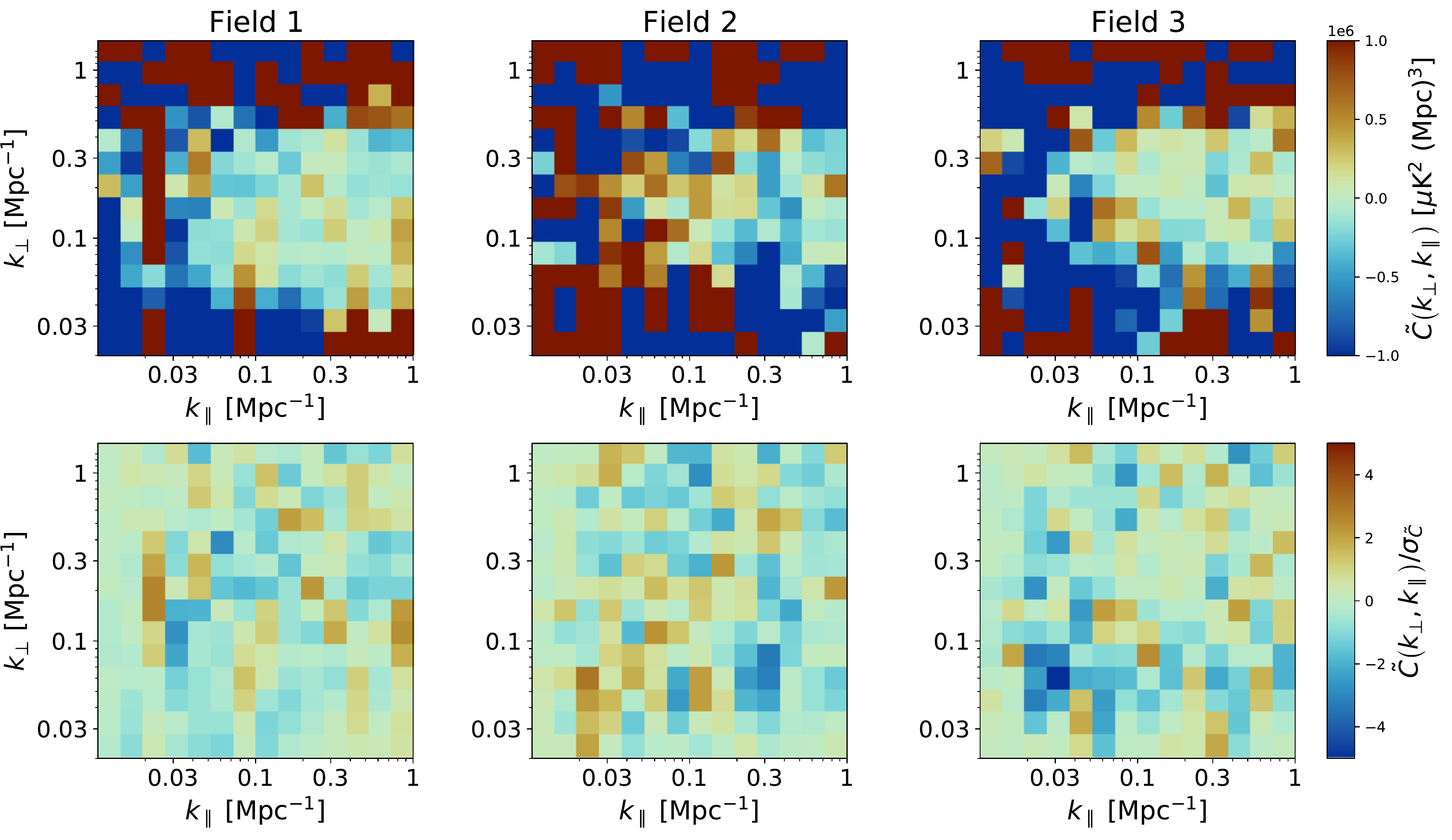}
		\caption{Cylindrically averaged mean pseudo-cross-spectra for CES observations (top row). Second row shows the spectra divided by the corresponding white noise uncertainty.\label{fig:ps_2d_CES}}
	\end{center}
\end{figure*}

Figure~\ref{fig:ps_2d_CES} shows the corresponding cylindrically averaged power spectrum in $k_\parallel \times k_\bot$ space. We see that the noise blows up very quickly as we move away from the region where the transfer function peaks. This illustrates the importance of taking the transfer function into account during the spherical averaging if you want maximum sensitivity in the 1D spectrum. In the region where we have appreciable sensitivity the spectra look consistent with white noise, as for the 1D spectra. The bottom row shows the spectra divided by their corresponding white noise uncertainty, to illustrate better what happens at all scales.

\begin{figure*}[t]
	\begin{center}
		\includegraphics[width=1.0\textwidth]{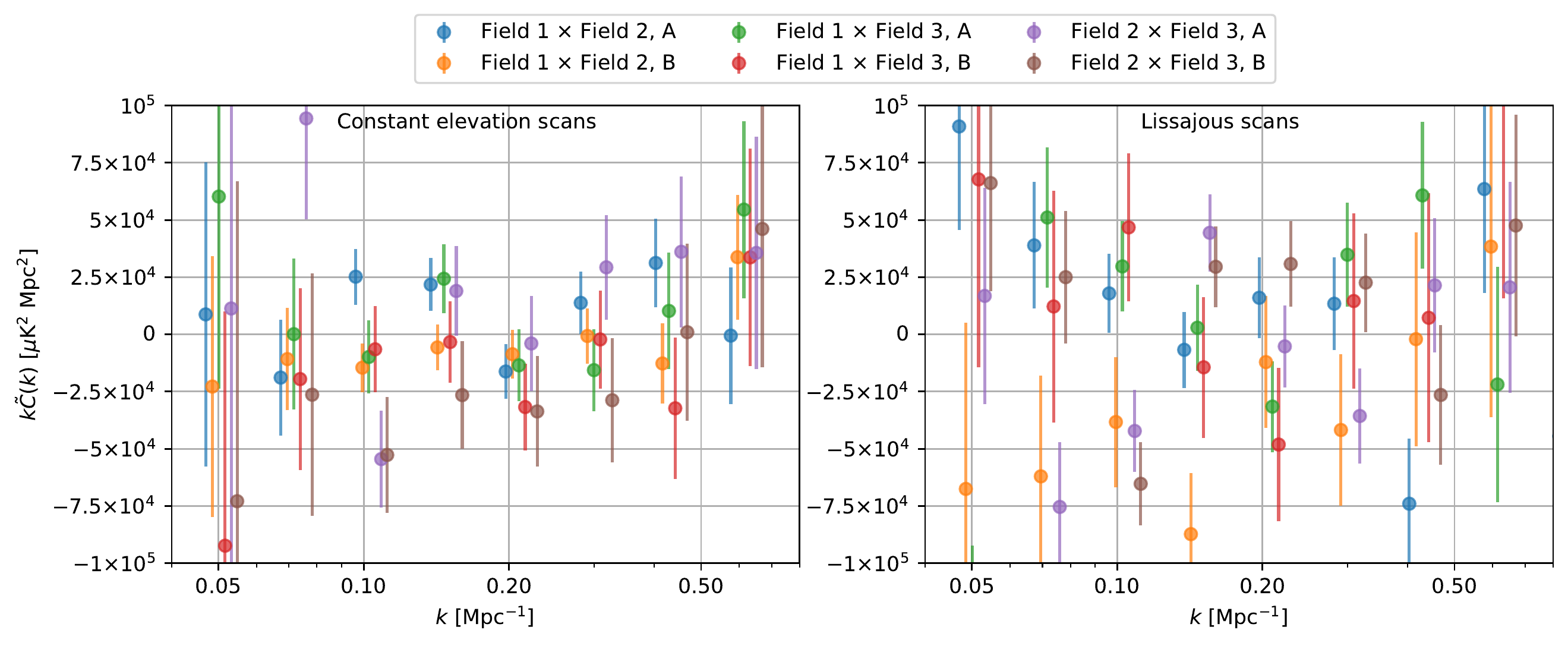}
		\caption{FPXS spectra of maps from different fields. Here A denote cross-spectra of the low elevation map from the first and the high elevation map from the second field, while B denotes the opposite combination. Data points from the different spectra are offset slightly in $k$ from their actual values to make them easier to distinguish.\label{fig:null}}
	\end{center}
\end{figure*}

\subsection{Null tests}
Given that the current data appear to be largely consistent with white noise, the importance of null-tests is less critical than if a potential detection had been made. Still, null-tests are a useful consistency check, and may be useful to identify and highlight specific systematic errors, and they may potentially provide hints regarding the nature of the Lissajous excesses. 

In order to get a sensitive set of null-tests, we can calculate cross-spectra between maps of our different fields. In these null-tests, any systematic related to standing waves or from residual large scale ground contamination could still show up, while the signal should not contribute at all. Moreover, as long as we center the fields appropriately, they are roughly as sensitive as our original spectra, while other null-tests are typically less sensitive. We therefore perform the same kind of FPXS power spectrum estimation as for the main results, but use the high elevation data from one field and the low elevation data from another field. In this way we obtain two null-tests per field pair, one where the first field uses the low elevation data while the second field uses the high elevation data (denoted A), and another pair (denoted B) when the first field uses the high elevation data and the other the low elevation data. This gives a total of 6 null tests for each scanning method. 

Figure~\ref{fig:null} shows the results from these calculations, and we see that the null spectra are consistent with white noise expectations for all the CES data. For the Lissajous data, however, we see that most of the null-tests show large excesses in power, consistent with our interpretation of the main Lissajous data containing systematics. 

Table~\ref{tab:null} shows $\chi^2$ statistics calculated from each of the null-tests in Figure~\ref{fig:null} as well as the single field results from Figures~\ref{fig:FPXS_1d_CES} and \ref{fig:FPXS_1d_liss} by combining the 8 datapoints of each spectrum. Specifically we caluclate the ``Probability to exceed", defined as one minus the cumulative distribution function (CDF) of the $\chi^2$ distribution with the given number of degrees of freedom (here eight). If the data was given by white noise, these statistics should be evenly distributed between 0 and 100~\%, while if we have excess power present, then the values will tend to be small. The results in the table supports and quantifies the statements we made above, that the CES data looks consistent with white noise and that the Lissajous data has clear power excesses present.

Although it is hard to interpret precisely, the fact that we see the clear excess in most of the Lissajous spectra made by combining maps from different fields, suggests that whatever systematic effect gives rise to this excess, it needs to be common to all the fields. 

\begin{table*}[t]
	\caption{$\chi^2$ statistics from science results (left) and null-tests using all the different Field combinations (right). These were calculated by combining the datapoints shown in Figures~\ref{fig:FPXS_1d_CES}, \ref{fig:FPXS_1d_liss} and \ref{fig:null} into a single $\chi^2$ value for each spectrum. \label{tab:null}}
	\begin{center}
		\begin{tabular}{r | c | c | c | c || c | c | c | c | c |  c}
			${ }$ & \multicolumn{8}{c}{$\chi^2$, Probability to exceed (PTE)} \\
			\hline
			Fields & All & Field 1 & Field 2 & Field 3 & $1 \times 2$, A & $1 \times 2$, B & $1 \times 3$, A & $1 \times 3$, B & $2 \times 3$, A & $2 \times 3$, B  \\
			\hline
			CES & 33 \% & 17 \% & 30 \% & 52 \% & 9 \% & 73 \% & 52 \% & 69 \% & 5 \% & 28 \% \\
			Lissajous & 0.02 \% & 0.1 \% & 3 \% & 72 \% & 3 \% & 3 \% & 0.5 \% & 58 \%  & 0.3 \% & 0.3 \% \\
		\end{tabular}
	\end{center}
\end{table*}

\section{Conclusion}
\label{sec:discussion}

In this paper, we have introduced the FPXS as a robust method for estimating the CO signal power spectrum from 3D intensity maps produced by the COMAP data analysis pipeline. We have discussed how to account for signal loss due to both filtering and beam smoothing, and we have estimated their magnitudes for the first-year COMAP observations with simulations. Computing the FPXS from the actual COMAP data, we find that the current data set is consistent with white noise for constant elevation scan data, and the uncertainties average down with time as expected for ideal data. Equivalently, these results suggest that all systematic errors are lower than the white noise level in our main sensitivity range. 

In contrast, the FPXS results from the Lissajous scan data show clear signs of systematic errors. Further modelling and analysis work is required before these data can be used for astrophysical analysis.

Null tests largely seem consistent with our main results, with all CES null tests being consistent with white noise, while most of the Lissajous null-tests show clear excesses, supporting our assumptions that the excesses seen in the main Lissajous spectra are the result of systematics. 

Future analysis will involve explicit estimation of the mode mixing matrix (see Appendix~\ref{app:mode_mixing}) to undo the mode mixing effect and present unbiased power spectrum estimates. We can also increase our sensitivity (by up to a factor of $\sqrt{2}$) by splitting the data into more than the current low and high elevation pieces. 

\begin{acknowledgements}
This material is based upon work supported by the National Science Foundation under Grant Nos.\ 1517108, 1517288, 1517598, 1518282 and 1910999, and by the Keck Institute for Space Studies under ``The First Billion Years: A Technical Development Program for Spectral Line Observations''. Parts of the work were carried out at the Jet Propulsion Laboratory, California Institute of Technology, under a contract with the National Aeronautics and Space Administration, and funded through the internal Research and Technology Development program. 

DTC is supported by a CITA/Dunlap Institute postdoctoral fellowship. The Dunlap Institute is funded through an endowment established by the David Dunlap family and the University of Toronto. CD acknowledges support from an STFC Consolidated Grant (ST/P000649/1). 

JB, HKE, MKF, HTI, JGSL, MR, NOS, DW, and IKW acknowledge support from the Research Council of Norway through grants 251328 and 274990, and from the European Research Council (ERC) under the Horizon 2020 Research and Innovation Program (Grant agreement No.\ 819478, \textsc{Cosmoglobe}). 
JG acknowledges support from the University of Miami and is grateful to Hugh Medrano for assistance with cryostat design. 
SH acknowledges support from an STFC Consolidated Grant (ST/P000649/1). 
J.\ Kim is supported by a Robert A.\ Millikan Fellowship from Caltech. 
At JPL, we are grateful to Mary Soria for for assembly work on the amplifier modules and to Jose Velasco, Ezra Long and Jim Bowen for the use of their amplifier test facilities. 
HP acknowledges support from the Swiss National Science Foundation through Ambizione Grant PZ00P2{\_}179934. PCB is supported by the James Arthur Postdoctoral Fellowship. We thank Isu Ravi for her contributions to the warm electronics and antenna drive characterization. The Scientific color maps \texttt{roma} and \texttt{tokyo}~\citep{Crameri2018} are used in this study to prevent visual distortion of the data and exclusion of readers with color-vision deficiencies~\citep{Crameri2020}.

Finally, we want to thank the anonymous referee, whose comments and suggestions have helped to improve and clarify this manuscript.
\end{acknowledgements}
\software{Matplotlib~\citep{matplotlib}; Astropy, a community-developed core Python package for astronomy~\citep{astropy}.}

\appendix

\begin{figure}[t]
	\begin{center}
		\includegraphics[width=0.49\textwidth]{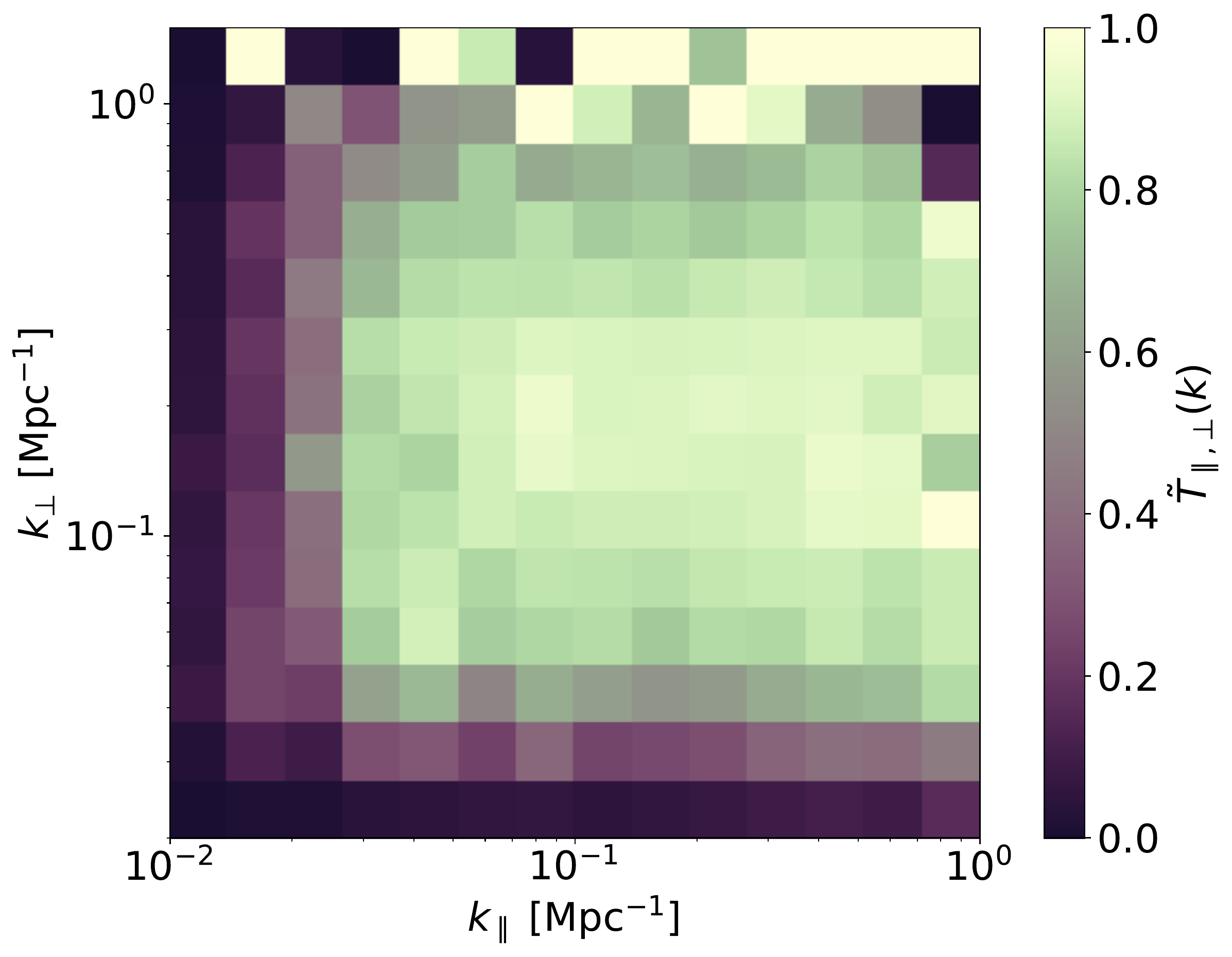}
		\vskip -1em
		\caption{Pipeline transfer functions for the cylindrically averaged power spectrum for Lissajous scans. This transfer function is based on a single signal realization and roughly three hours of data. \label{fig:TF_2d_liss}}
	\end{center}
\end{figure}

\section{Lissajous power spectrum results}\label{app:liss}
When looking at the Lissajous data, we found clear excess power in the spectra. For this reason, we do not use any Lissajous data for our final science results, but include the Lissajous power spectra here for completeness. We use the exact same approach for the Lissajous data as we did for the CES data. 
We derive a separate pipeline transfer function for the Lissajous data (see Figure~\ref{fig:TF_2d_liss}), since the properties of these scans are a bit different from the CES's. As noted in \cite{es_III} we find that the transfer function for the Lissajous data preserves a bit more large scale angular structure than the CES one, but it does not make much qualitative difference in terms of the complete transfer function.  
In the automatic 5$\sigma$ cut based on the $\chi^2_{C(\vec k)}$ statistics, the fraction of remaining spectra that were removed for the Lissajous data was given by 132/182 for Field 1, 92/210 for Field 2 and 109/210 for Field 3. 

\begin{figure}[t]
	\begin{center}
		\includegraphics[width=0.49\textwidth]{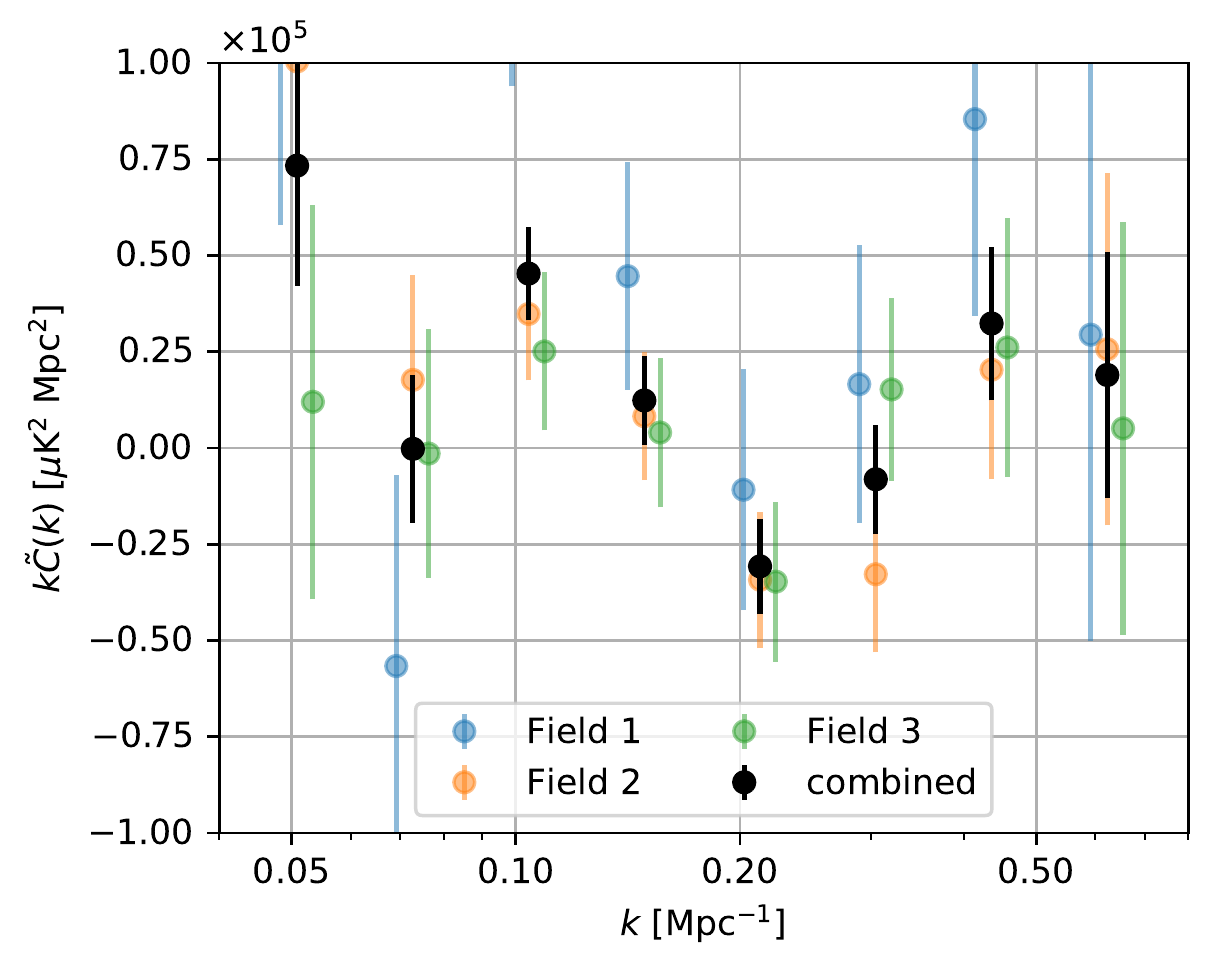}
		\caption{Spherically averaged mean pseudo-cross-spectra for Lissajous observations of Field 1 (blue), Field 2 (orange) and Field 3 (green). These spectra were generated from all the accepted data using the FPXS cross-spectrum statistic. In addition the full transfer function has been applied, to de-bias the signal estimate. Data points from the different fields are offset slightly in $k$ from their actual values to make them easier to distinguish. \label{fig:FPXS_1d_liss}}
	\end{center}
\end{figure}

The Lissajous FPXS results for the spherically averaged power spectrum are shown in Figure~\ref{fig:FPXS_1d_liss}. 
In contrast to the CES data, these data do not appear equally well behaved. Here, we see clear signs of excess power on large scales in both the Field 1 and Field 2. These excesses suggest that large scale systematic errors are still present in for the Lissajous scans, and may for instance be caused by either residual atmospheric variations or ground pickup from the far sidelobes. 

\begin{figure*}[t]
	\begin{center}
		\includegraphics[width=1.0\textwidth]{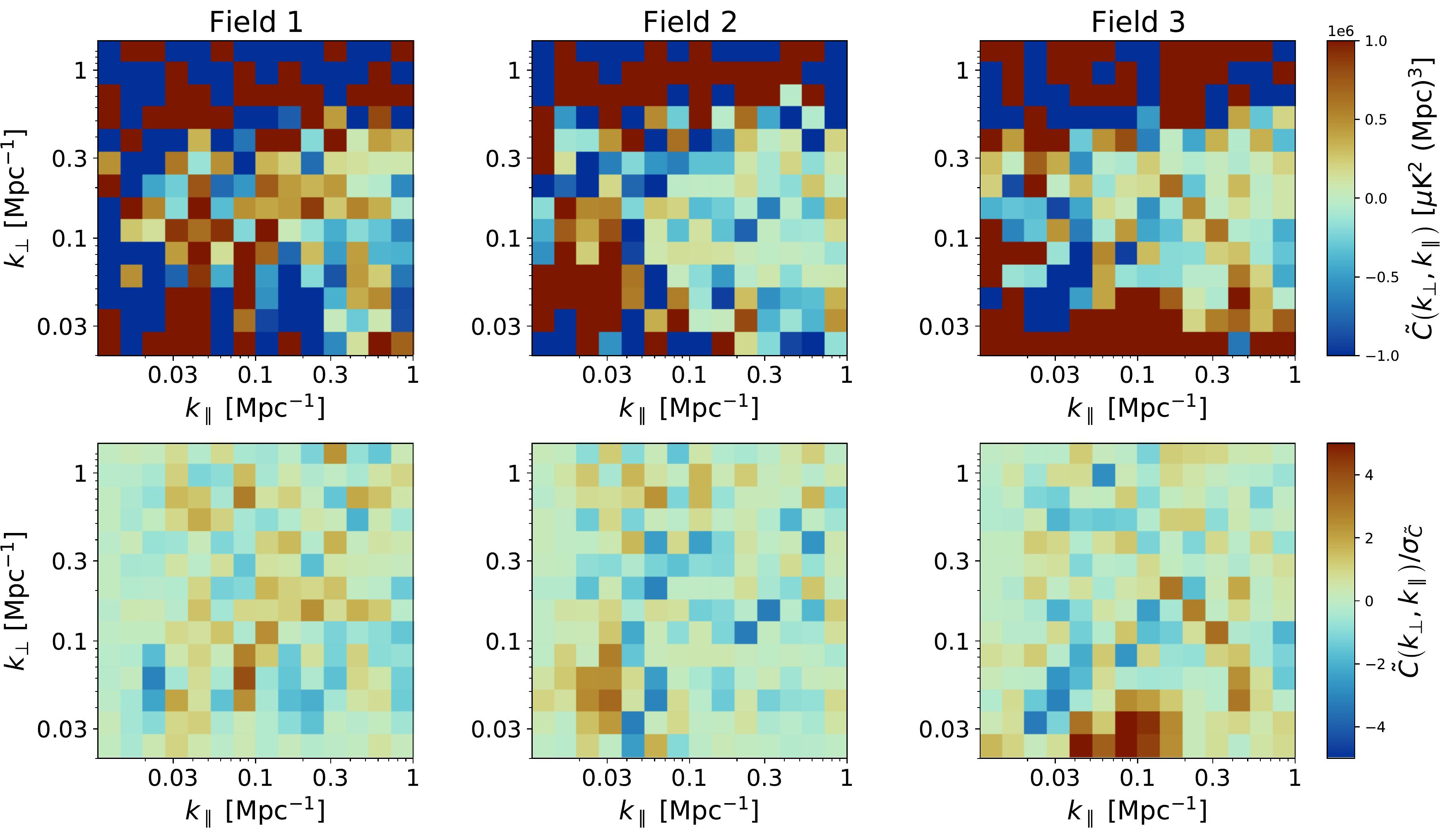}
		\caption{Cylindrically averaged mean pseudo-cross-spectra for Lissajous observations (top row). Second row shows the spectra divided by the corresponding white noise uncertainty.\label{fig:ps_2d_liss}}
	\end{center}
\end{figure*}

These residuals are even more prominent when considering the 2D $k_\parallel \times k_\bot$ power spectrum, as shown in Figure~\ref{fig:ps_2d_liss}. Here we see some clear regions exhibiting systematic power excess. This is seen most clearly in the second row of the figure, which shows the power spectrum divided by the expected white noise fluctuations, and thus correspond to power measured in units of standard deviations. In particular, for Field 3 we see a bright region on the largest angular scales, and on scales between $k_\parallel \sim 0.03$--0.1 Mpc${}^{-1}$ in the frequency direction. We also see a fairly bright region at around $k_\bot = 0.2$ Mpc${}^{-1}$ and between $k_\parallel \sim 0.06$--0.6 Mpc${}^{-1}$in the Field 1 data, which is right in the middle of our most sensitive region. 

\section{Fourier conventions}
In this Appendix, we present the conventions for the discrete Fourier transformations used in this paper. All the conventions are consistent with the default conventions in NumPy's \citep{numpy} FFT library. The forward transformation is given by
$$f_l = \sum_{m=0}^{n-1} x_m \exp\left( - 2\pi i \frac{ml}{n}\right), \ \ \ \ l = 0, \cdots, n-1$$ 
where $x_m$ are the discrete values of the function in real space, and $f_l$ are the Fourier coefficients. The inverse transformation is then given by
$$x_m = \frac{1}{n} \sum_{l=0}^{n-1} f_l \exp\left( 2\pi i \frac{ml}{n}\right),$$
We define the physical wave number 
\begin{align*}
 k &\equiv \frac{2 \pi j}{\Delta x n}, \ \ j \in \left\{ -\frac{n}{2}, \cdots, -1, 0, 1, \cdots, \frac{n}{2} \right\} \\
 	&= 2\pi \cdot \text{np.fft.fftfreq}(n, \Delta x).
\end{align*}

\section{Definition of cosmological map grid}
Since Fourier transforms require a rectangular grid, we assume that the 3D temperature maps can be approximated by a rectangular grid in co-moving cosmological parameters. We assume that all the voxels have the same shape and size as the middle voxel at redshift $z_\textrm{mid} \approx 2.9$.

The comoving length corresponding to an angular separation $\delta \theta$, for a given redshift $z$, is given by
\begin{equation}
 \delta l_\bot = r(z) \delta \theta   = \delta \theta \int_0^z \frac{c dz'}{H(z')},
\end{equation}
where $r(z)$ is the comoving distance travelled by light emitted from redshift $z$ to us. 

The comoving radial distance corresponding to a small redshift interval $\delta z = z_1- z_2 = \nu_0/\nu_1^\text{obs} - \nu_0/\nu_2^\text{obs} \approx (1+z)^2 \delta \nu^\text{obs} /\nu_0$, where $z_1 > z_2$, is given by
\begin{equation}
 \delta l_\parallel = \int_{z_2}^{z_1} \frac{c dz}{H(z)} \approx \frac{c \delta z}{H(z)} \approx \frac{c}{H(z)} \frac{(1+z)^2 \delta \nu^\text{obs}}{\nu_0}, 
\end{equation}
where $\nu_0 \approx 115.27$ is the emission frequency of the CO 1$\rightarrow$0 line we are studying and $\delta \nu^\text{obs}$ = 31.25 MHz is the resolution of our frequency bins. 

Given a pixel width of 2 arcmin, we then get the following voxel dimensions
\begin{align}
    \delta l_\bot &= 3.63 \mathrm{Mpc}, \\
    \delta l_\parallel &= 4.26 \mathrm{Mpc}. 
\end{align}

\section{Mode mixing and the master algorithm}\label{app:mode_mixing}
In order to understand the mode-mixing effect, let us consider in more detail the Fourier transform of a weighted map\footnote{We work in 2D here to save some indices; the generalization to 3D is straightforward.},
\begin{equation}
	 \tilde{f}_{k_1 k_2} = \sum_{m_1=0}^{n-1}\sum_{m_2=0}^{n-1} x_{m_1 m_2} W_{m_1 m_2}\exp\left( - 2\pi i \frac{m_1k_1 + m_2k_2}{n}\right).
\end{equation}
Here $x_{m_1 m_2}$ is the map, $W_{m_1 m_2}$ is the weight map and $\tilde{f}_{k_1 k_2}$ is the Fourier transform of the weighted map. We can insert the expression for the inverse Fourier transform of $x$ and $W$, 
\begin{equation}
	 \tilde{f}_{k_1 k_2} = \frac{1}{n^4} \sum_{k^{'}_1=0}^{n-1}\sum_{k^{'}_2=0}^{n-1} f_{k^{'}_1 k^{'}_2} \sum_{k^{''}_1=0}^{n-1}\sum_{k^{''}_2=0}^{n-1} f^W_{k^{''}_1 k^{''}_2}  \sum_{m_1=0}^{n-1}\sum_{m_2=0}^{n-1}
	 \exp\left( - 2\pi i \frac{m_1(k^{'}_1 + k^{''}_1 - k_1) + m_2(k^{'}_2 + k^{''}_2 - k_2)}{n}\right),
\end{equation}
where $f$ and $f^W$ are the Fourier transforms of $x$ and $W$ respectively. Working through the algebra, we get
\begin{align}
 \tilde{f}_{k_1 k_2} &= \frac{1}{n^2} \sum_{k^{'}_1=0}^{n-1}\sum_{k^{'}_2=0}^{n-1} f_{k^{'}_1 k^{'}_2} \sum_{k^{''}_1=0}^{n-1}\sum_{k^{''}_2=0}^{n-1} f^W_{k^{''}_1 k^{''}_2}  \delta_{k^{''}_1(k_1 - k^{'}_1)\%n, k^{''}_2(k_2 - k^{'}_2)\%n}\nonumber\\
 \tilde{f}_{k_1 k_2} &= \frac{1}{n^2} \sum_{k^{'}_1=0}^{n-1}\sum_{k^{'}_2=0}^{n-1} f_{k^{'}_1 k^{'}_2} f^W_{(k_1 - k^{'}_1)\%n (k_2 - k^{'}_2)\%n} \nonumber\\
 \tilde{f}_{k_1 k_2} &= \sum_{k^{'}_1=0}^{n-1}\sum_{k^{'}_2=0}^{n-1} f_{k^{'}_1 k^{'}_2} \underbrace{\frac{1}{n^2} f^W_{(k_1 - k^{'}_1)\%n (k_2 - k^{'}_2)\%n}}_{\equiv \, K_{k_1, k_2, k^{'}_1, k^{'}_2}},
\end{align}
where $\%$ denotes the modulo operation and where we have defined the mode mixing amplitude $K_{\vec{k},\vec{k'}}$. 

Adopting vector notation, we may now write the pseudo-spectrum as follows,
\begin{align}
	\tilde P(\vec{k}) &= \frac{V_\mathrm{vox}}{N_\mathrm{vox}} \langle \tilde{f}_{\vec{k}}\tilde{f}_{\vec{k}}^*\rangle \\
			&= \frac{V_\mathrm{vox}}{N_\mathrm{vox}} \frac{1}{n^{2D}} \sum_{\vec{k'}} \sum_{\vec{k}^{''}} \langle f_{\vec{k'}}  f^*_{\vec{k}^{''}} \rangle K_{\vec{k},\vec{k'}} K^*_{\vec{k},\vec{k}^{''}} \\
			&=\frac{1}{n^{2D}} \sum_{\vec{k'}} \sum_{\vec{k{''}}} P(\vec{k'}) \delta_{\vec{k}',\vec{k}^{''}}K_{\vec{k},\vec{k}'} K^*_{\vec{k},\vec{k}^{''}} \\
			&= \sum_{\vec{k'}} P(\vec{k'}) \underbrace{\frac{1}{n^{2D}} |K_{\vec{k},\vec{k}'}|^2}_{M_{{\vec{k},\vec{k}'}}},
\end{align}
where $D$ is the number of dimensions of the map, and where we have defined the mode mixing matrix, $M_{{\vec{k},\vec{k}'}}$. We see that the auto-spectrum and the pseudo-spectrum is related by a linear transformation, so all the information in one is also there in the other. 

Within the CMB field, accounting for mode mixing by explicitly calculating and inverting $M_{{\vec{k},\vec{k}'}}$ is often referred to as the MASTER algorithm \citep{hivon2002,Leung2021}. Doing this requires that we calculate the mode mixing between each Fourier mode and all the other Fourier modes, so for a 3D maps this scales poorly with the map dimension. On the other hand, the algorithm parallelizes trivially, and the matrix must only be computed once for a given weight map, after which the same operation may be applied efficiently to any number of simulations. Whether this is feasible depends on the details of the individual use case. Some methods exists in the literature to approximate this procedure in a faster way, see e.g. \cite{Louis2020}.

\bibliography{bibfile,bibfile_thesis,early_science}{}
\bibliographystyle{aasjournal}

\end{document}